\documentclass[11pt]{article}

\usepackage[preprint]{acl}

\usepackage{times}
\usepackage{latexsym}

\usepackage[T1]{fontenc}

\usepackage[utf8]{inputenc}

\usepackage{microtype}

\usepackage{inconsolata}

\usepackage{graphicx}
\usepackage{booktabs}
\usepackage{textcomp}   
\usepackage{amssymb}    
\usepackage{tabularx}
\usepackage[table]{xcolor}
\usepackage{tcolorbox}
\tcbuselibrary{breakable,skins,listings}
\usepackage{enumitem}
\usepackage{multirow}

\newcommand{\benchmark}{FinanceComplexQA}
\newcommand{\skill}{Finance-LaTeX}

\newcolumntype{Y}{>{\raggedright\arraybackslash}X}
\definecolor{PromptFrame}{HTML}{2C3E50}
\definecolor{PromptFill}{HTML}{F8FAFC}
\definecolor{PromptAccent}{HTML}{3B6EA5}
\definecolor{PromptMuted}{HTML}{64748B}
\definecolor{NoteFill}{HTML}{FFF8E7}
\definecolor{NoteFrame}{HTML}{D4A017}

\newtcolorbox{promptmeta}[1]{
  enhanced,
  breakable,
  colback=PromptFill,
  colframe=PromptAccent,
  arc=2.5mm,
  boxrule=0.6pt,
  left=10pt,
  right=10pt,
  top=8pt,
  bottom=8pt,
  title={#1},
  fonttitle=\bfseries\sffamily,
  coltitle=white,
  attach boxed title to top left={yshift=-2.5mm, xshift=5mm},
  boxed title style={
    colback=PromptAccent,
    arc=1.5mm,
    boxrule=0pt,
    top=2pt,
    bottom=2pt,
    left=6pt,
    right=6pt
  }
}

\newtcolorbox{promptnote}{
  enhanced,
  breakable,
  colback=NoteFill,
  colframe=NoteFrame,
  arc=2mm,
  boxrule=0.5pt,
  left=10pt,
  right=10pt,
  top=6pt,
  bottom=6pt,
  fontupper=\small
}

\NewTCBListing{promptbody}{ !O{} }{
  enhanced,
  breakable,
  colback=PromptFill,
  colframe=PromptFrame,
  arc=2mm,
  boxrule=0.4pt,
  left=8pt,
  right=8pt,
  top=6pt,
  bottom=6pt,
  listing only,
  listing options={
    basicstyle=\small\ttfamily,
    breaklines=true,
    breakatwhitespace=false,
    columns=fullflexible,
    keepspaces=true,
    showstringspaces=false,
    aboveskip=0pt,
    belowskip=0pt
  },
  #1
}

\newcommand{\placeholder}[1]{\texttt{\textless #1\textgreater}}

\title{\benchmark{}: Benchmarking Agentic Reasoning on Industrial-grade Financial Documents}

\author{
  {\bf Xianfu Cheng}\textsuperscript{\rm 1},
  {\bf Shiwei Zhang}\textsuperscript{\rm 2},
  {\bf Jiyu Zhao}\textsuperscript{\rm 3},
  {\bf Jian Yang}\textsuperscript{\rm 1\dag},  
  {\bf Xinyuan Wang}\textsuperscript{\rm 3},  
  {\bf Ming Zhou}\textsuperscript{\rm 4}\\
  {\bf Weixiao Zhou}\textsuperscript{\rm 1},
  {\bf Xiangyuan Guan}\textsuperscript{\rm 1},
  {\bf Xiang Li}\textsuperscript{\rm 1},   
  {\bf Zhenhe Wu}\textsuperscript{\rm 1}, 
  {\bf Ziyi Ni}\textsuperscript{\rm 3},
  {\bf Zhoujun Li}\textsuperscript{\rm 1,5},
  {\bf Bingjing Xu}\textsuperscript{\rm 2}\\
  \textsuperscript{\rm 1}Beihang University; \textsuperscript{\rm 2}Microsoft,  China;
  \textsuperscript{\rm 3}Multilingual-Multimodal-NLP;\\ 
  \textsuperscript{\rm 4}Langboat Technology, Beijing, China; \textsuperscript{\rm 5}Shenzhen Intelligent Strong Technology Co.,Ltd.\\
  \texttt{\{buaacxf,jiayang,lizj\}@buaa.edu.cn,shiweizhang@microsoft.com}\\
}

\begin{document}
\maketitle
\begin{abstract}
Agentic Reasoning has become a transformative force in financial analysis due to its ability to integrate large-scale information and generate reliable and accurate content. However, when handling complex real-world problems, different agents still show significant performance variation.
In this work, we design {\bf Finance-LaTeX} SKILL, a skill for synthesizing financial documents with complex layouts based on expert knowledge. Using an agent workflow built on this skill, we generate 2,000 professional financial documents along with 6,000 high-quality question-answer pairs.
To evaluate the overall capability of agents, we introduce {\bf FinanceComplexQA}, a comprehensive open-ended generation benchmark for financial documents that closely resembles real-world scenarios. It contains 2,026 deep research tasks targeting 1009 financial documents. FinanceComplexQA has 8 key features: bilingual support; coverage of six mainstream scenarios and seven tasks; expert-level document reasoning questions; deep research of complex layouts; relatively stable and permanent reference answers; and precise evaluation through an Agent-as-a-Judge with multiple evaluation metrics.
Using FinanceComplexQA, we conduct a comprehensive evaluation of leading RAG systems and agentic reasoning tools for financial document QA. Through identifying and analyzing failure cases, we provide an in-depth study of their capabilities in numerical computation, multi-hop reasoning, content summarization, and industry analysis.

\end{abstract}

\section{Introduction}

In recent years, large language models (LLMs)~\cite{deepseek_v3,llama,qwen3,yang2025code,yang2026iquest,loopcoderv2} have achieved remarkable progress in natural language understanding, reasoning, and multimodal generation, enabling their adoption in high-stakes domains such as finance, law, and healthcare. Among these, finance presents unique challenges due to its strong requirements for numerical precision, domain expertise, and factual reliability. Unlike open-domain question answering, financial analysis often requires multi-step reasoning over long documents, integration of structured and unstructured information, and strict consistency with objective evidence. Even minor factual or reasoning errors may lead to flawed investment decisions or risk assessments, making finance one of the most demanding scenarios for reliable LLM deployment.

Recent advances such as retrieval-augmented generation (RAG) systems~\cite{lewis2020retrieval,guo2024lightrag,xiang2025use,xiao2025graphrag}, Model Context Protocol (MCP) tools~\cite{schick2023toolformer,li2025review,liu2026dive}, and Agentic Reasoning tools~\cite{anthropic2026claudecode,openai2026harness} have improved LLM performance on knowledge-intensive tasks by enabling external retrieval, dynamic tool invocation, and multi-stage planning. However, evaluating such systems in real financial scenarios remains difficult. Existing financial question-answering (QA) benchmarks mainly focus on narrow tasks such as table QA~\cite{wu2025tablebench,wu2026mmtablebench,shu2026m3tqa}, numerical extraction~\cite{cheng2025xformparser,wu2025breaking}, short-context fact retrieval~\cite{wei2024measuring,he2025chinese,cheng2025simplevqa}, or Summarization~\cite{zhou2023multi,zhou2025they,zhou2026large}. They rarely capture the complex reasoning process required in real financial research, where analysts must synthesize evidence across documents, tables, charts, and domain knowledge. Moreover, many existing benchmarks rely on synthetic questions or template-based construction, limiting realism and reasoning depth. Current evaluation protocols also struggle to provide stable and consistent scoring for open-ended analytical answers.

To address these limitations, we introduce {\bf FinanceComplexQA}, a large-scale benchmark for open-ended complex question answering over financial documents. FinanceComplexQA contains 2026 expert-level questions built from 1009 real-world financial rich-text documents, covering 8 financial subdomains and 9 representative financial research tasks, including numerical analysis, multi-hop reasoning, causal inference, industry comparison, and cross-layout synthesis. The benchmark is designed to evaluate whether modern LLM-based systems can perform document understanding, deep reasoning, and analytical synthesis at a level closer to that of professional financial analysts.

Before the benchmark, we propose {\bf Finance-LaTeX} SKILL, a multi-agent framework for scalable financial QA data synthesis. The framework combines expert knowledge, automated document generation, terminology constraints, and financial logic verification to generate high-quality financial documents and question-answer pairs. Using this framework, we synthesize an additional 2,000 financial documents and 6,000 QA samples, providing a scalable and automated path for benchmark construction and future benchmark expansion.
FinanceComplexQA is built around four core design principles. First, we introduce a dual-context reasoning paradigm, where solving each question requires both explicit document evidence and implicit financial domain knowledge. Models must retrieve relevant evidence and combine it with financial logic and industry knowledge for accurate reasoning. Second, we emphasize cross-layout reasoning: each question requires a joint understanding of textual passages, tables, forms, and contents, forcing systems to aggregate evidence across heterogeneous document elements. Third, the benchmark covers diverse financial tasks and scenarios, enabling fine-grained evaluation of different reasoning capabilities. Fourth, to ensure long-term usability, all questions are grounded in relatively stable and verifiable financial facts and are evaluated with a unified framework that jointly measures semantic correctness, numerical accuracy, and reasoning completeness through an Agent-as-a-Judge pipeline.

We evaluate 2 mainstream RAG/MCP systems and 2 agentic reasoning systems on FinanceComplexQA. Results reveal a substantial performance gap between current state-of-the-art systems and real-world financial reasoning requirements. Even advanced systems struggle with long-chain numerical reasoning, cross-layout evidence fusion, and industry-level analytical synthesis. Detailed error analysis further exposes critical weaknesses in reasoning, planning, evidence utilization, and factual consistency, highlighting key directions for future research in reliable financial AI agents.

Our contributions are summarized as follows:
(1) We propose Finance-LaTeX SKILL, a multi-agent framework for scalable generation of high-quality financial documents with QA data;
(2) We present FinanceComplexQA, a large-scale benchmark for complex financial document question answering that covers diverse tasks, domains, and multi-step reasoning scenarios;
(3) We design a challenging benchmark construction methodology centered on dual-context reasoning and cross-layout evidence aggregation, enabling more realistic evaluation of financial analytical reasoning;
(4) We conduct a systematic evaluation of state-of-the-art RAG, MCP, and agentic systems, revealing major bottlenecks in complex financial reasoning and providing insights for future financial LLM research.

\begin{table*}[!t]
    \centering
    \small
    \resizebox{\textwidth}{!}{
    \begin{tabular}{lcccccccc} 
        \toprule
         \textbf{Benchmark} & \textbf{LNG} & \textbf{\#Docs} &\textbf{\#QAs} & \textbf{PDF parser} & \textbf{Long doc} & \textbf{Cross-Layout} & \textbf{Open-ended} & \textbf{Metrics} \\
        \midrule
        FinQA~\cite{chen2021finqa} & EN & 2789 pages & 1147 & \texttimes & \texttimes & \texttimes & \checkmark & Recall, EA, PA \\ 
        ConvFinQA~\cite{chen2022convfinqa} & EN & 434 pages & 1568 & \texttimes & \texttimes & \checkmark & \texttimes & EA, PA \\
        DocFinQA~\cite{reddy2024docfinqa} & EN & 100 & 922 & \checkmark & \checkmark & \checkmark & \texttimes & HR, ACC \\
        TAT-QA~\cite{zhu2021tatqa} & EN & 275 context & 1655 & \texttimes & \texttimes & \checkmark & \texttimes & EM, F1 score \\
        FinanceBench~\cite{islam2023financebench} & EN & 150 context & 150 & \texttimes & \texttimes & \checkmark & \checkmark & CA, IA, F2A \\
        officeQA~\cite{databricks2024officeqa} & EN & 246 & 246 & \checkmark & \checkmark & \checkmark & \texttimes & ACC, ARE \\
        officeQA-Pro~\cite{databricks2025officeqapro} & EN & 696 & 133 & \checkmark & \checkmark & \texttimes & \texttimes & ACC, ARE \\
        BizFinBench~\cite{lu2025bizfinbench} & EN/CN & 6500 context & 6500 & \texttimes & \texttimes & \texttimes & \checkmark & ACC \\
        BizFinBench v2~\cite{guo2026bizfinbench} & EN/CN & 29578 context & 29578 & \texttimes & \checkmark & \texttimes & \checkmark  & ACC \\
        GDPval~\cite{patwardhan2025gdpval} & EN & 423 & 220 & \checkmark & \checkmark & \checkmark & \checkmark & WO, WT \\
        FinanceAgent~\cite{bigeard2025finance} & EN & tool env. & 200 & \texttimes & \texttimes & \texttimes & \checkmark & CBA, ACC \\
        \midrule
        \textbf{\benchmark{} (Ours)} & \textbf{EN/CN} & \textbf{1009} & \textbf{2026} & \textbf{\checkmark} & \textbf{\checkmark} & \textbf{\checkmark} & \textbf{\checkmark} & \textbf{ACC, ROU, FS, Cov} \\
        \bottomrule
    \end{tabular}}
    \caption{Comparison between \benchmark{} and representative financial and enterprise QA benchmarks in terms of language coverage, scale, PDF parsing, long-document support, cross-layout reasoning, open-ended answering, and evaluation metrics. Compared with prior datasets, \benchmark{} jointly targets bilingual expert questions over parsed financial documents and evaluates open-ended analytical answers with accuracy, overlap, faithfulness, and coverage metrics.}
    \label{tab: benchmark_compare}
\end{table*}

\section{\skill{} SKILL}

\subsection{Design Motivation}

\skill{} is designed as the document-generation engine behind the benchmark construction process illustrated in Figure~\ref{fig:bench_construction}. Its goal is to synthesize professional financial documents and question-answer pairs that resemble realistic analyst workloads rather than isolated text snippets. To achieve this, the skill combines expert financial knowledge, evidence planning, layout-aware LaTeX writing, and multi-stage verification. The generated data are used as a scalable development source for stress testing, prompt iteration, and benchmark expansion, while the held-out evaluation benchmark remains grounded in curated financial documents.

The skill is motivated by two gaps in existing financial QA data. First, many datasets contain short excerpts or simplified tables, while real financial analysis often requires reading long reports with paragraphs, tables, forms, captions, and special layouts. Second, open-ended answers require more than factual extraction: they must connect evidence to financial logic, preserve units and time periods, and avoid unsupported recommendations. \skill{} therefore treats a document as a structured evidence environment and makes every generated QA pair pass explicit consistency and solvability checks.

\subsection{Expert Knowledge and Evidence Planning}

The workflow begins with corpus collection and expert knowledge acquisition. For each candidate document, an agent samples a financial domain, target audience, document type, terminology constraints, and reasoning targets. The selected domain concepts include stable accounting relations, regulatory terms, market-analysis concepts, and business-process facts that are unlikely to become obsolete quickly. This design follows the left part of Figure~\ref{fig:bench_construction}, where corpus evidence and expert knowledge are combined before document generation.

Before writing a document, the skill constructs an evidence plan. The plan specifies which facts should appear in paragraphs, which quantities should appear in tables, which assumptions should be expressed in captions or notes, and which pieces of evidence should later support questions. Evidence is also classified by reasoning function, such as direct lookup, numerical comparison, multi-hop inference, summary evidence, or planning constraint. This makes the generated documents suitable for testing both explicit retrieval and implicit financial reasoning.

\subsection{Layout-Aware LaTeX Generation}

The generation stage converts the evidence plan into a LaTeX document with realistic financial structure. Instead of producing plain text, the agent writes sections, paragraphs, tables, special layout blocks, formulas, captions, and table-adjacent descriptions. Numeric cells are generated together with row labels, column labels, periods, units, and entity scopes, so that later questions can require the system to bind each number to its correct context.

The skill intentionally creates cross-layout dependencies. For example, a paragraph may describe a change in operating strategy, a table may give the corresponding revenue and margin values, and a note may define the scope of consolidation. A valid answer must connect these elements rather than relying on a single sentence. This layout-aware design is central to \benchmark{}, because Figure~\ref{fig:bench_construction} emphasizes special layouts and paragraph evidence as inputs to benchmark construction.

\subsection{QA Pair Generation and Verification}

After document generation, the workflow creates candidate question-answer pairs from the planned evidence. Each question is paired with required evidence units, a reference answer, and reasoning notes. The question generator rejects items that can be answered by copying a single span, that depend on volatile external facts, or that ask for unsupported speculation. The answer generator then writes analytical responses that include conclusions, evidence, calculations, and caveats.

\begin{figure*}[t]
\centering
\includegraphics[width=1.0\linewidth]{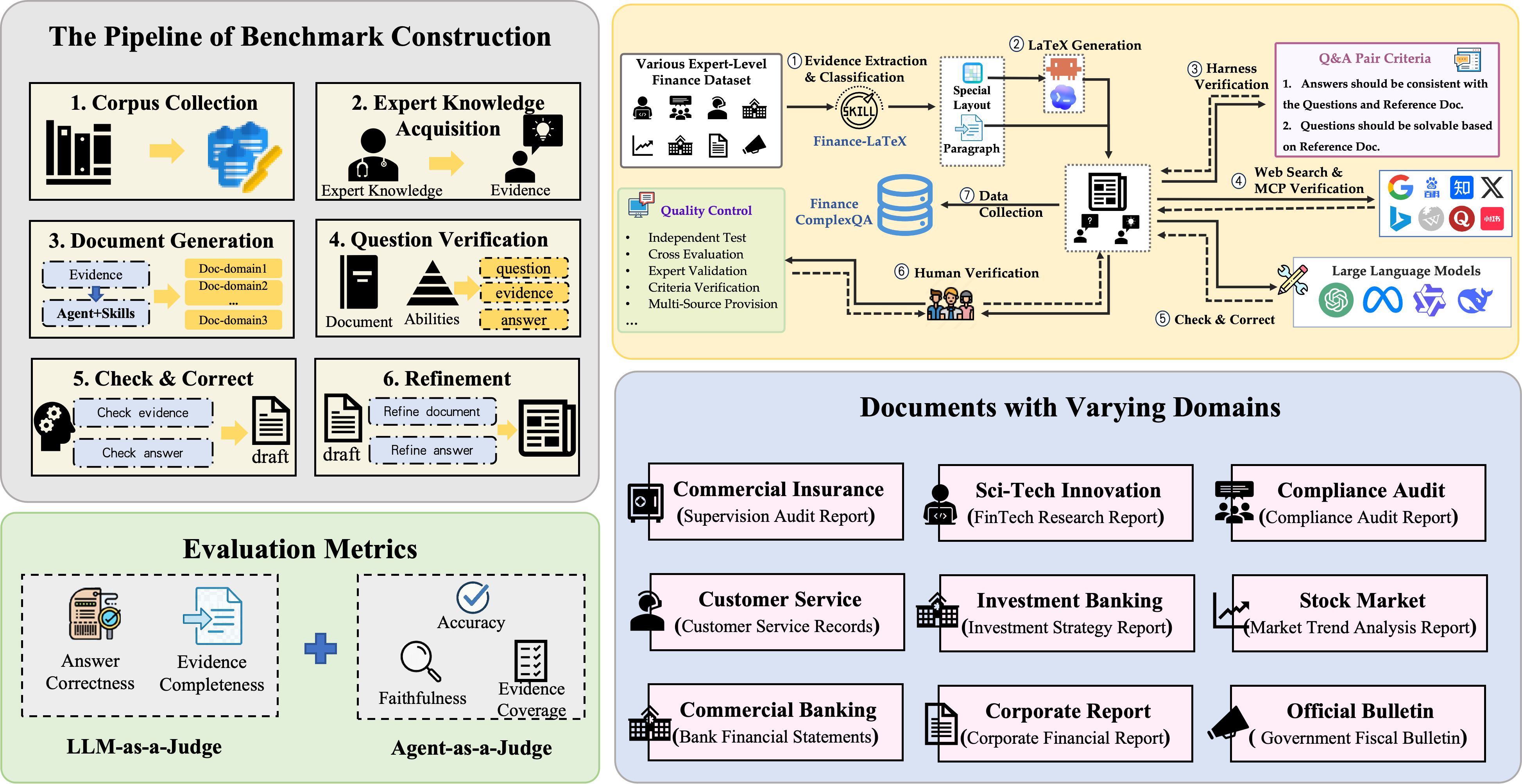}
\caption{Overview of the \benchmark{} construction pipeline. The workflow collects financial corpora, incorporates expert knowledge, generates layout-rich documents with \skill{}, verifies question-answer pairs through LLM, Web/MCP, and human checks, and organizes the resulting benchmark by financial domain and judge-oriented evaluation metrics.}
\label{fig:bench_construction}
\end{figure*}

The generated QA pairs are checked against the criteria shown in Figure~\ref{fig:bench_construction}: answers should be consistent with the questions and reference documents, and questions should be solvable from the reference documents. Verification is performed through multiple channels. LLM-based checks inspect coherence, completeness, and financial logic; web search and MCP-style tools verify stable public facts when needed; and harness tests check whether the evidence-answer relation is reproducible. If a document or answer fails these checks, the workflow returns to a check-and-correct stage before the sample can enter the data pool.

\subsection{Quality Control and Data Usage}

The final stage applies quality control at both document and QA levels. The checks include independent testing, cross evaluation, expert validation, criteria verification, and multi-source provision, matching the quality-control block in Figure~\ref{fig:bench_construction}. At the document level, we remove samples whose layout is broken, whose tables lose headers or units, or whose financial relations are inconsistent. At the QA level, we remove samples with ambiguous evidence, irreproducible calculations, incomplete reference answers, or weak reasoning depth.

Using this workflow, we generate 2{,}000 professional financial documents and 6{,}000 high-quality QA pairs. These synthetic data are not mixed into the held-out benchmark for reporting model performance. Instead, they support development and ablation studies, provide controlled cases for testing retrieval and agent behavior, and help expose failure modes before systems are evaluated on \benchmark{}.

\section{\benchmark{} Pipeline}

\subsection{Overview}

\benchmark{} is a bilingual benchmark for open-ended generation over industrial-grade financial documents. As summarized in Figure~\ref{fig:bench_construction}, the pipeline combines real financial corpora, expert evidence extraction, \skill{}-assisted document and QA construction, LLM/web/MCP verification, human review, and judge-oriented evaluation. The resulting benchmark contains 2{,}026 deep research tasks over 1009 financial documents, with 1{,}013 Chinese questions and 1{,}013 English questions.

The benchmark is designed to test whether an agent can perform the full reasoning loop expected in financial analysis: locate evidence, preserve layout context, apply domain knowledge, compute or compare quantities, synthesize an answer, and remain faithful to the source. This is why the pipeline records not only questions and reference answers, but also evidence units, reasoning notes, scenario labels, task labels, and evaluation dimensions.

\subsection{Corpus and Financial Scenarios}

\benchmark{} is constructed from professional financial documents rather than isolated article fragments. The corpus includes corporate financial reports, investment strategy reports, market trend analysis reports, FinTech research reports, bank financial statements, customer service records, supervision and compliance audit materials, and government fiscal bulletins. The metadata of these corpora are respectively derived from ConvFinQA~\cite{chen2022convfinqa}, FinanceBench-test~\cite{islam2023financebench}, officeQA-Pro~\cite{databricks2025officeqapro}, BizFinBench v2~\cite{guo2026bizfinbench} (Counterfactual Inference, Anomaly Information Tracing), and the actual industrial financial knowledge base.

The financial domains shown in Figure~\ref{fig:bench_construction}, such as commercial banking, corporate reports, investment banking, stock-market analysis, customer service, compliance audit, commercial insurance, Sci-Tech innovation, and official bulletins, are normalized into the scenario labels used in Table~\ref{tab:bench_statistics}. These scenarios were selected because they require different evidence habits. Corporate reports emphasize accounting relations; investment reports require market and industry synthesis; compliance documents require careful rule interpretation; customer service records stress event reconstruction and intent understanding; and fiscal bulletins require linking policy descriptions to amounts, periods, and program structure.

The documents are parsed into layout-aware units, including paragraphs, tables, forms, captions, section titles, and table-adjacent descriptions. We retain this structural information because losing page and layout context often removes the decisive evidence required by financial reasoning.

\begin{table*}[!t]
\centering
\resizebox{\linewidth}{!}{
\begin{tabular}{lrr|lrr}
\toprule
\multicolumn{2}{l}{\multirow{2}*{\textbf{Statistics}}} & \textbf{Count} & \multicolumn{2}{l}{\multirow{2}*{\textbf{Statistics}}} & \textbf{Count} \\

 & \textbf{\#QAs} & \textbf{\#Docs} &  & \textbf{\#QAs} & \textbf{\#Docs} \\
\hline
\textbf{Data} & 2026 & 1009 & \textbf{8 types of Scene Categories} &  &   \\
- Chinsne (CN) & 1013 & 625 & - FinTech Research Report (FR), CN\&EN & 134+157 & 65+25  \\
- English (EN) & 1013 & 384 & - Corporate Financial Report (CF), CN\&EN & 163+253 & 141+58 \\
\textbf{9 types of Task Categories} &  &  & - Investment Strategy Report (IS), CN\&EN  & 246+276 & 138+136 \\
- Planning (PLA), CN\&EN & 137+101 & 74+61 & - Market Trend Analysis Report (MTA), CN\&EN & 91+192 & 43+30 \\
- Summary (SUM), CN\&EN & 154+243 & 154+159 &- Bank Financial Statements (BFS), CN & 139 & 85 \\
- Implicit Reasoning (IR), CN\&EN & 163+220 & 141+76 & - Customer Service Records (CSR), CN & 140 & 80 \\
- Comparison (CMP), CN & 190 & 119 & - Supervision and Compliance Audit (SCA), CN & 100 & 73 \\
- Knowledge Query (KQ), CN & 141 & 81 & - Government Fiscal Bulletin (GFB), EN & 135 & 135 \\
- MultiHop Reasoning (MR), CN & 228 & 158 & \textbf{(CN/EN) Words length of } & \textbf{Question} & \textbf{Golden} \\

- Explicit Reasoning (ER), EN & 195 & 167 & - Maximum length & 255/1141 & 5671/2623 \\
- MultiHop Judgment (MJ), EN & 141 & 76 & - Minimum length & 5/51 & 5/1 \\
- Numerical Comparison (NC), EN & 113 & 47 & - Average length & 63/286 & 283/366 \\
\bottomrule
\end{tabular}
}
\caption{Dataset statistics of \benchmark{}, including the bilingual split, QA and document counts by task category and financial scene, and question/reference-answer length distributions. Counts joined by ``+'' denote Chinese and English subsets, respectively.}
\label{tab:bench_statistics}
\end{table*}

\begin{figure*}[t]
\centering
\includegraphics[width=1.0\linewidth]{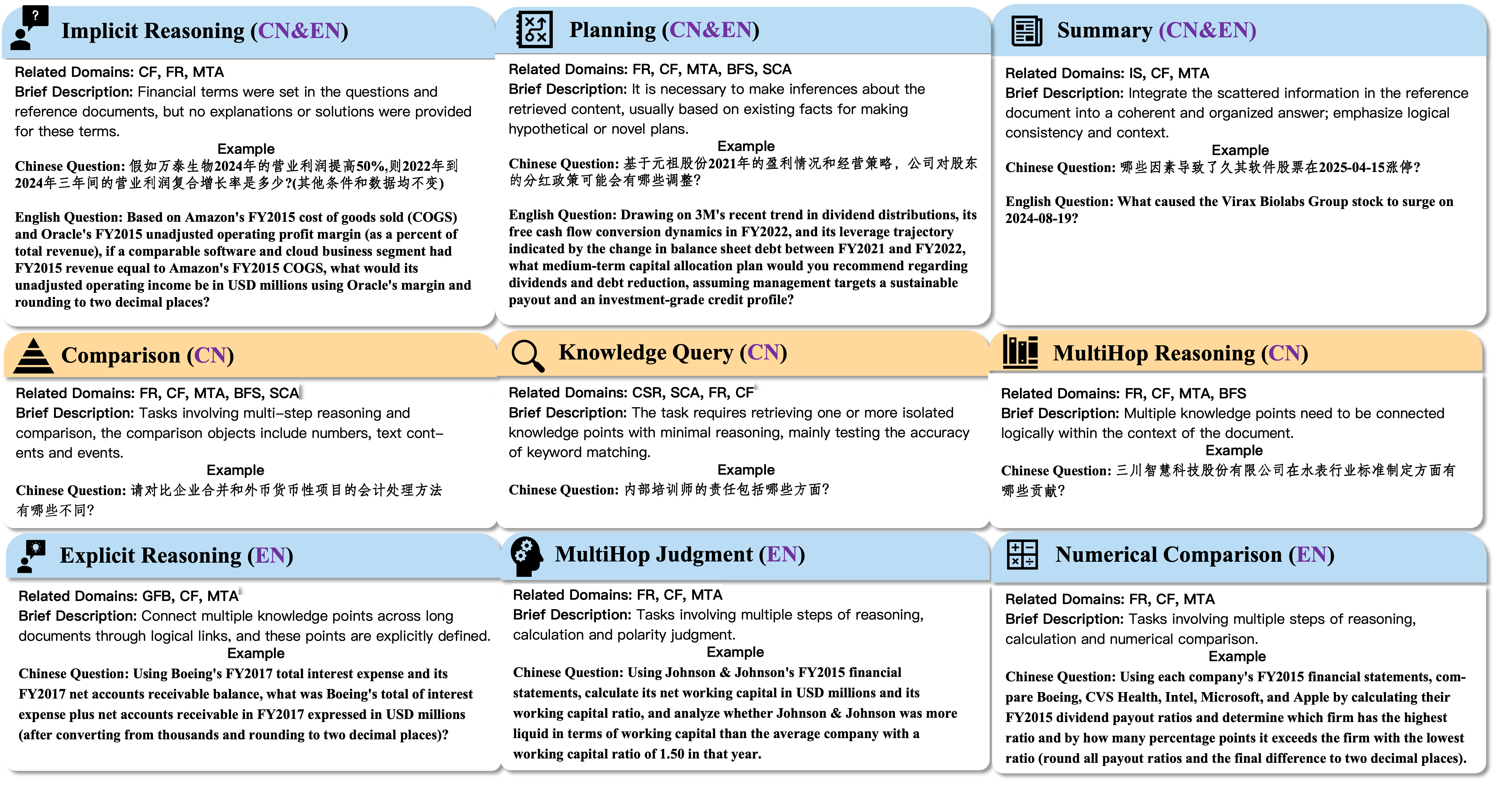}
\caption{Task taxonomy and representative benchmark cases. The figure summarizes the nine Chinese and English task labels, their related financial domains, brief reasoning requirements, and example questions, ranging from implicit reasoning and planning to multi-hop judgment and numerical comparison.}
\label{fig:task_benchcases}
\end{figure*}

\begin{table}[t]
\centering
\small
\setlength{\tabcolsep}{8pt}
\begin{tabularx}{\linewidth}{cccc}
\toprule
\textbf{LNG} & \textbf{Group} & \textbf{Pages (avg.)} & \textbf{Length (avg.)}\\
\midrule
 & FR & 18.23 & 12127\\
 & CF & 9.43 & 7359\\
 & IS & \textbf{22.33} & 15191\\
 & MTA & 21.47 & \textbf{15803}\\
 & BFS & 19.87 & 12276\\
 & CSR & 19.90 & 11606\\
  & SCA & 19.99 & 12736\\
\multirow{-8}{*}{CN} & \textbf{AVG} & 18.01 & 12040\\
\midrule
 & FR & 17.72 & 30911\\
  & CF & 14.78 & 24943\\
   & IS & 14.23 & 27519\\
    & MTA & \textbf{21.30} & \textbf{33498}\\
 & GFB & 9.92 & 16978\\
\multirow{-6}{*}{EN} & \textbf{AVG} & 13.58 & 24112\\
\midrule
\multicolumn{2}{c}{\textbf{Total}}  & 16.33 & 16493\\
\bottomrule
\end{tabularx}
\caption{Average document length by language and financial scene in \benchmark{}. The table reports mean page counts and mean token/word lengths for each Chinese and English scenario group, showing that the benchmark is dominated by long, layout-rich financial documents.}
\label{tab:document_statistics}
\end{table}


\subsection{Task Design}

The benchmark questions are written to represent real analytical requests rather than template slots. We use nine task labels. Chinese questions include comparison (CMP), implicit reasoning (IR), knowledge query (KQ), multi-hop reasoning (MR), summarization (SUM), and planning (PLA). English questions include numerical comparison (NC), implicit reasoning (IR), multi-hop judgment (MJ), explicit reasoning (ER), summarization (SUM), and planning (PLA). The labels are not meant to isolate mutually exclusive cognitive skills. Instead, they provide a primary lens for evaluation and error analysis.

Each question is checked against three design requirements. First, it must have a stable reference answer based on relatively persistent financial facts or document-internal evidence. Second, it must require more than one trivial retrieval step. Third, it must be answerable by a careful analyst using the supplied documents and common financial knowledge. Questions that depend on live market prices, rapidly changing policies, or unsupported speculation are removed or rewritten.

\subsection{Document Parsing and Evidence Units}

Each document is converted into a hierarchy of evidence units before question writing. The hierarchy preserves page order, section boundaries, paragraph spans, table row and column headers, captions, and nearby explanatory text. We treat tables as structured evidence rather than plain strings: numeric cells are stored together with their row labels, column labels, units, and period information. This representation is important for financial reasoning because the same number can have different meanings depending on whether it refers to a quarter, a fiscal year, a consolidated entity, or a business segment.

For long documents, evidence units are grouped into page-local blocks and document-level topic blocks. Page-local blocks support layout-sensitive retrieval, while topic blocks support cross-page reasoning. During annotation, question writers can mark which units are necessary, optional, or misleading. This lets us distinguish a system that retrieves the right page from one that actually uses the decisive evidence.

\subsection{Annotation Schema}

For each example, we record the question text, language, scenario label, task label, required evidence units, reference answer, and reasoning notes. The reasoning notes include intermediate quantities, comparison targets, and domain assumptions when they are needed to reproduce the answer. For cross-layout questions, the schema also records the evidence relation, such as text-to-table, table-to-chart, or multi-page synthesis. This metadata is used only for evaluation and analysis; models do not receive the gold evidence relation at inference time.

The schema supports fine-grained error diagnosis. If a model retrieves all required evidence but reaches the wrong conclusion, the failure is classified as a reasoning or calculation error. If it misses a table, chart, or footnote that the reference depends on, the failure is classified as evidence omission. This distinction is useful because retrieval failures and reasoning failures require different system improvements.

\subsection{Dual-Context and Cross-Layout Reasoning}

The most important property of \benchmark{} is dual-context reasoning. In ordinary document QA, the relevant sentence may contain most of the answer. In financial research, a sentence usually becomes meaningful only after it is interpreted through accounting definitions, market structure, risk logic, or regulatory background. For example, a table may show that revenue grew while operating cash flow declined; the answer requires not only extracting both numbers, but also explaining why the divergence matters.

We therefore design questions that combine explicit and implicit contexts. Explicit context includes document spans, table cells, chart labels, and stated assumptions. Implicit context includes domain knowledge such as margin interpretation, debt pressure, sector cyclicality, accounting relations, or the difference between nominal growth and real operating improvement. The reference answer records both types of evidence, allowing evaluators to penalize answers that are fluent but incomplete.

Cross-layout reasoning is enforced by construction. A typical question may require retrieving a management discussion paragraph, matching it to a financial table, and using a chart or trend statement to determine whether the conclusion is supported. This is difficult for systems that flatten documents into independent chunks. It is also difficult for agents that retrieve evidence correctly but fail to preserve the relationship between a row header, a time column, and a surrounding narrative explanation.

\subsection{Reference Answers and Evaluation}

Reference answers are written as analytical responses rather than minimal spans. They include the conclusion, the key evidence, relevant calculations, and caveats. For numerical answers, the reference records the formula or comparison logic whenever possible. For planning and summary tasks, the reference identifies mandatory coverage points so that a shorter answer can still be credited if it includes the essential reasoning.

We evaluate model outputs with an Agent-as-a-Judge protocol. The judge receives the question, reference answer, retrieved evidence where available, and the candidate answer. It assigns task-specific scores for accuracy, semantic alignment, numerical correctness, evidence coverage, faithfulness, and completeness. For planning and summarization tasks, coverage and faithfulness receive higher weight. For numerical comparison and explicit reasoning, exact quantities and units receive higher weight. We also record cost and latency, because deployment in financial workflows is constrained by both quality and resource use.

\subsection{Quality Control}

The benchmark is filtered through both automatic and expert-oriented checks. At the document level, we remove files whose parsing output loses section hierarchy, breaks table headers, or produces ambiguous chart descriptions. At the question level, we reject items that can be answered by copying a single span, that depend on transient market data, or that ask for subjective investment recommendations without document support. At the answer level, we check whether mandatory evidence points are present and whether numeric operations are reproducible from the cited values.

We also maintain a difficulty balance. If a subset contains too many direct lookup questions, additional multi-hop or cross-layout questions are sampled from the same scenario. Conversely, if a question requires private external knowledge or speculative assumptions, it is rewritten to expose enough evidence in the reference documents. This keeps the benchmark difficult while preserving answerability. The goal is not to make every example maximally hard, but to reflect the range of tasks that a financial analyst would face in a realistic research workflow.

\section{Experiments}

\subsection{Systems}



We evaluate systems that represent three common ways of using LLMs for financial document QA. The first is a lightweight retrieval baseline, represented by LightRAG~\cite{guo2024lightrag}. The second is a layout-aware retrieval pipeline, represented by PageIndex~\cite{zhang2025pageindex}, which is designed to preserve page-level and table-adjacent structure. The third is an agentic setting, represented by Codex-style and Claude Code-style agents~\cite{openai2024codex,anthropic2024claudecode}, where the system can plan, call tools, inspect intermediate evidence, and revise an answer before final generation.

We evaluate 7 representative LLMs in total, comprising 4 closed-source and 3 open-source models, which provide a diverse evaluation of model capabilities across different architectures and training paradigms.
The closed-source models include GPT-5.4 and GPT-5.5\footnote{\url{https://platform.openai.com/docs/models\#gpt-5.4;\#gpt-5.5}}, Claude-5-Sonnet\footnote{\url{https://www.anthropic.com/news/claude-sonnet-5}}, Qwen3.7-plus\footnote{\url{https://dashscope.aliyuncs.com\#/model-market}}.
The open-source models cover 2 frameworks, including Qwen3.5-flash and Qwen3.5-plus\footnote{\url{https://huggingface.co/Qwen\#Qwen3.5-35B-A3B;\#Qwen3.5-397B-A17B}}, DeepSeek-v4-flash\footnote{\url{https://platform.deepseek.com/}}.

Each system is paired with either closed-source or open-source foundation models, as shown in Tables~\ref{tab:scene_results}--\ref{tab:aggregate_results}. All systems receive the same document collection and question set. We keep each system's native retrieval or agent loop, but require a comparable final answer format. For every answer, we record the question language, scenario, task label, judge scores, token usage, and latency. This design lets us compare overall quality while also diagnosing whether a system fails because of retrieval, calculation, answer synthesis, or cost.

\begin{table}[t]
\centering
\small
\setlength{\tabcolsep}{3.5pt}
\begin{tabularx}{\linewidth}{lY}
\toprule
\textbf{Dimension} & \textbf{Judgment criterion} \\
\midrule
Accuracy & The final conclusion matches the reference answer. \\
Numeric correctness & Calculations preserve formula, unit, period, and sign. \\
Evidence coverage & Required text, table, and chart evidence is used. \\
Faithfulness & Claims are supported by the provided documents. \\
Completeness & The answer includes required caveats and reasoning steps. \\
Cost & Token usage and tool calls remain practical for deployment. \\
\bottomrule
\end{tabularx}
\caption{Evaluation dimensions used by the Agent-as-a-Judge protocol.}
\label{tab:evaluation_dimensions}
\end{table}

\subsection{Metrics}



Table~\ref{tab:evaluation_dimensions} summarizes the Agent-as-a-Judge protocol. Accuracy measures whether the conclusion matches the reference answer. Numeric correctness checks whether formulas, units, periods, signs, and entity scopes are preserved. Evidence coverage measures whether required paragraphs, tables, charts, and captions are used. Faithfulness penalizes claims that are not supported by the provided documents. Completeness captures whether the answer includes required caveats and reasoning steps. Cost records whether the token usage and tool calls are practical for deployment.

The task tables use metric combinations that match the answer type. Accuracy (ACC) is used across tasks as the primary correctness signal. ROUGE-style overlap (ROU) is reported for retrieval and reasoning tasks as a lexical-alignment diagnostic, but it is not treated as sufficient for correctness. Coverage (Cov) is emphasized for summarization and planning, where an answer may be fluent but omit required evidence. Faithfulness (FS) is reported for planning tasks, where generic or unsupported recommendations are a major risk. The aggregate averages in Table~\ref{tab:aggregate_results} are computed over accuracy-style scores, while the detailed task table preserves the metric-level differences needed for diagnosis.

\begin{table*}[!t]
\resizebox{1.\textwidth}{!}{
\begin{tabular}{c|c|cccccccc|cccccc}
\toprule
\multirow{2}{*}{\textbf{Agent}} & \multirow{2}{*}{\textbf{Core Model}} & \multicolumn{8}{c|}{\textbf{Chinese Financial Scenes}} & \multicolumn{6}{c}{\textbf{English Financial Scenes}} \\ & & \textbf{FR} & \textbf{CF} & \textbf{IS} & \textbf{MTA} & \textbf{BFS} & \textbf{CSR} & \textbf{SCA} & \textbf{OA} & \textbf{FR} & \textbf{CF} & \textbf{IS} & \textbf{MTA} & \textbf{GFB} & \textbf{OA}\\
\midrule
\multicolumn{16}{c}{\cellcolor[HTML]{EFEFEF}\textit{Closed-Source Large Language Models}} \\
LightRAG & gpt-5.5 & 80.59 & 59.20 & 41.46 & 55.49 & 77.85 & 66.07 & 74.50 & 62.42 & 57.64 & 76.87 & 53.80 & 61.19 & 41.48 & 59.92 \\
\midrule
PageIndex & gpt-5.4 & 86.92 & 57.05 & 64.78 & 63.41 & 85.56 & 70.15 & 85.35 & 72.15 & 67.62 & \underline{\textbf{83.33}} & 65.58 & 67.12 & 37.69 & 66.90\\
\midrule

 & gpt-5.5 & 79.83 & \underline{72.12} & 63.87 & 55.15 & 72.37 & 68.93 & 76.59 & 69.29 & \underline{\textbf{70.38}} & 76.48 & 64.49 & 64.58 & \underline{\textbf{70.37}} & 69.20\\
\multirow{-2}{*}{Codex} & qwen3.7-plus & 88.05 & 69.32 & 65.04 & 65.38 & 84.39 & 72.14 & 85.50 & 74.48 & 64.33 & 77.57 & 66.30 & 65.36 & 62.22 & 68.08\\
\midrule
 & sonnet 5 & \underline{\textbf{91.04}} & 65.64 & \underline{\textbf{68.69}} & \underline{\textbf{67.03}} & \underline{\textbf{87.41}} & \underline{\textbf{73.21}} & 87.00 & \underline{\textbf{76.01}} & 65.60 & 80.63 & 65.94 & \underline{\textbf{67.96}} & 61.85 & \underline{\textbf{69.39}}\\
\multirow{-2}{*}{Claude Code} & qwen3.7-plus & 89.83 & 69.62 & 67.06 & 66.87 & 85.65 & 72.30 & \underline{\textbf{89.13}} & 75.93 & 68.14 & 75.42 & \underline{\textbf{67.05}} & 64.08 & 62.40 & 68.08\\
\midrule

\multicolumn{16}{c}{\cellcolor[HTML]{EFEFEF}\textit{Open-Source Large Language Models}} \\
 & qwen3.5-flash & 81.71 & 55.21 & 63.61 & 62.63 & 76.97 & 65.35 & 79.50 & 68.21 & 63.69 & 73.71 & 55.61 & 62.50 & 49.25 & 61.84 \\
  & qwen3.5-plus & 81.34 & 52.45 & 59.34 & 63.73 & 80.21 & 63.21 & 74.00 & 66.38 & 65.28 & 75.09 & 34.05 & 63.54 & 47.40 & 56.51\\
\multirow{-3}{*}{Codex} & deepseek-v4-flash & 86.94 & 72.69 & 64.17 & 60.43 & \underline{83.81} & 71.07 & 80.80 & \underline{74.66} & 69.74 & 76.87 & \underline{64.80} & 63.80 & \underline{66.79} & \underline{69.33}\\
\midrule
& qwen3.5-flash & 88.72 & 54.43 & \underline{65.44} & \underline{64.60} & 83.21 & \underline{72.62} & 82.32 & 71.82 & 64.10 & 77.09 & 57.42 & 63.87 & 48.86 & 63.58\\
& qwen3.5-plus & \underline{88.80} & 68.75 & 62.22 & 61.18 & \underline{83.81} & 71.42 & 82.77 & 73.24 & 68.58 & 76.20 & 51.63 & 65.64 & 59.30 & 64.49 \\
\multirow{-2}{*}{Claude Code} & deepseek-v4-flash & 85.82 & \underline{\textbf{73.14}} & 57.11 & 61.79 & 82.01 & 67.66 & \underline{85.00} & 71.58 & \underline{\textbf{70.38}} & \underline{77.86} & 56.02 & \underline{66.40} & 61.36 & 66.40\\
\bottomrule
\end{tabular}
}
\caption{Scenario-level answer-evaluation results judged by GPT-5-mini. Scores are reported for RAG, page-indexed, and agentic systems across Chinese and English financial scenes; OA denotes the overall average within each language block, and underlined values highlight the strongest results in the corresponding scenario.}
\label{tab:scene_results}
\end{table*}


\begin{table*}[!t]
\resizebox{1.\textwidth}{!}{
\begin{tabular}{c|c|cc|cc|cc|cc|cc|ccc}
\toprule
\multicolumn{15}{c}{\textbf{Chinese Financial Tasks}}\\
\midrule
\multirow{2}{*}{\textbf{Agent}} & \multirow{2}{*}{\textbf{Core Model}} & \multicolumn{2}{c|}{\textbf{CMP}} & \multicolumn{2}{c|}{\textbf{IR}} & \multicolumn{2}{c|}{\textbf{KQ}} & \multicolumn{2}{c|}{\textbf{MR}} & \multicolumn{2}{c|}{\textbf{SUM}} & \multicolumn{3}{c}{\textbf{PLA}} \\ & & \textbf{ACC} & \textbf{ROU} & \textbf{ACC} & \textbf{ROU} & \textbf{ACC} & \textbf{ROU} & \textbf{ACC} & \textbf{ROU} & \textbf{ACC} & \textbf{Cov} & \textbf{ACC} & \textbf{FS} & \textbf{Cov} \\
\midrule
\multicolumn{15}{c}{\cellcolor[HTML]{EFEFEF}\textit{Closed-Source Large Language Models}} \\

\multirow{-1}{*}{LightRAG} & gpt-5.5 & 55.86 & 23.46 & 54.67 & 42.72 & 58.20 & 32.88 & 56.11 & 19.17 & 44.87 & 31.93 & 51.43 & 58.36 & 49.57 \\
\midrule
\multirow{-1}{*}{PageIndex} & gpt-5.4 & 56.63 & 26.47 & 54.29 & 56.41 & \underline{\textbf{63.19}} & 35.77 & \underline{\textbf{65.19}} & \underline{27.93} & 58.36 & 49.43 & \underline{\textbf{56.53}} & 67.71 & 61.56 \\
\midrule

 & gpt-5.5 & \underline{56.66} & \underline{27.24} & \underline{73.33} & \underline{\textbf{71.17}} & 62.09 & \underline{\textbf{37.80}} & 61.02 & 27.00 & 60.08 & 47.58 & 50.58 & 83.54 & 32.77\\
\multirow{-2}{*}{Codex} & qwen3.7-plus & 55.95 & 23.09 & 69.63 & 70.24 & 61.67 & 32.96 & 58.47 & 24.44 & 60.03 & 57.61 & 54.15 & 92.31 & 64.90 \\
\midrule
 & sonnet 5 & 55.53 & 22.42 & 61.10 & 58.36 & 62.63 & 32.43 & 55.73 & 16.92 & \underline{60.29} & \underline{\textbf{60.57}} & 53.71 & \underline{\textbf{96.18}} & \underline{\textbf{68.36}}\\
\multirow{-2}{*}{Claude Code} & qwen3.7-plus & 50.42 & 20.06 & 60.84 & 55.99 & 60.22 & 31.38 & 55.55 & 22.92 & 51.87 & 45.63 & 51.92 & 84.59 & 60.85 \\
\midrule

\multicolumn{15}{c}{\cellcolor[HTML]{EFEFEF}\textit{Open-Source Large Language Models}} \\
 & qwen3.5-flash & 52.99 & 22.99 & 55.48 & 57.23 & 57.74 & 31.68 & 56.40 & 25.03 & 54.52 & 46.78 & 52.58 & 87.19 & 60.43 \\
 & qwen3.5-plus & 55.94 & 24.73 & 52.73 & 54.23 & 56.79 & 29.21 & 61.05 & \underline{\textbf{28.31}} & 46.85 & 28.33 & 50.61 & 90.20 & 51.76 \\
\multirow{-3}{*}{Codex} & deepseek-v4-flash & \underline{\textbf{57.10}} & \underline{\textbf{27.71}} & \underline{\textbf{73.37}} & \underline{70.61} & \underline{62.95} & \underline{35.04} & \underline{62.87} & 27.19 & 51.82 & 41.14 & \underline{54.55} & 83.96 & 55.81\\
\midrule
 & qwen3.5-flash & 52.91 & 21.34 & 48.53 & 41.19 & 59.49 & 31.53 & 57.59 & 24.80 & \underline{\textbf{61.17}} & \underline{58.86} & 53.88 & 87.40 & \underline{62.44} \\
 & qwen3.5-plus & 53.38 & 23.70 & 63.09 & 54.62 & 62.58 & 32.91 & 58.72 & 21.52 & 58.63 & 52.84 & 46.49 & 79.81 & 52.03 \\ 
\multirow{-3}{*}{Claude Code} & deepseek-v4-flash & 53.29 & 22.85 & 72.04 & 70.23 & 56.97 & 28.85 & 58.78 & 23.65 & 52.34 & 41.99 & 51.79 & \underline{90.92} & 61.40 \\
\midrule
\multicolumn{15}{c}{\textbf{English Financial Tasks}}\\
\midrule
\multirow{2}{*}{\textbf{Agent}} & \multirow{2}{*}{\textbf{Foundation Model}} & \multicolumn{2}{c|}{\textbf{NC}} & \multicolumn{2}{c|}{\textbf{IR}} & \multicolumn{2}{c|}{\textbf{MJ}} & \multicolumn{2}{c|}{\textbf{ER}} & \multicolumn{2}{c|}{\textbf{SUM}} & \multicolumn{3}{c}{\textbf{PLA}} \\ & & \textbf{ACC} & \textbf{ROU} & \textbf{ACC} & \textbf{ROU} & \textbf{ACC} & \textbf{ROU} & \textbf{ACC} & \textbf{ROU} & \textbf{ACC} & \textbf{Cov} & \textbf{ACC} & \textbf{FS} & \textbf{Cov} \\
\midrule
\multicolumn{15}{c}{\cellcolor[HTML]{EFEFEF}\textit{Closed-Source Large Language Models}} \\

\multirow{-1}{*}{LightRAG} & gpt-5.5 & 58.16 & 38.52 & 43.71 & 22.60 & 54.62 & 28.34 & 50.63 & 26.12 & 50.55 & 51.01 & 62.09 & 63.83 & 66.63 \\
\midrule
\multirow{-1}{*}{PageIndex} & gpt-5.4 & \underline{\textbf{61.12}} & 44.54 & 56.30 & 35.33 & \underline{\textbf{62.69}} & 35.90 & 54.51 & 36.35 & 55.95 & 56.29 & \underline{\textbf{71.35}} & 73.09 & 72.48 \\
\midrule

 & gpt-5.5 & 60.34 & \underline{\textbf{46.83}} & \underline{\textbf{59.64}} & \underline{\textbf{36.55}} & 62.57 & \underline{\textbf{39.59}} & \underline{\textbf{70.21}} & 56.82 & \underline{58.25} & 52.51 & 66.82 & 86.41 & 65.06 \\
\multirow{-2}{*}{Codex} & qwen3.7-plus & 55.61 & 38.96 & 59.33 & 31.87 & 56.80 & 34.37 & 68.60 & \underline{\textbf{57.64}} & 55.86 & \underline{\textbf{61.30}} & 63.00 & 92.31 & 73.15\\
\midrule

 & sonnet 5 & 52.94 & 35.66 & 51.76 & 27.41 & 54.53 & 29.92 & 59.78 & 37.63 & 55.07 & 55.16 & 60.98 & \underline{\textbf{92.76}} & \underline{74.74}\\
\multirow{-2}{*}{Claude Code} & qwen3.7-plus & 52.11 & 35.34 & 52.97 & 29.13 & 51.35 & 29.19 & 61.44 & 49.41 & 52.11 & 57.79 & 57.23 & 86.72 & 66.77 \\
\midrule

\multicolumn{15}{c}{\cellcolor[HTML]{EFEFEF}\textit{Open-Source Large Language Models}} \\

 & qwen3.5-flash & 53.17 & 35.43 & 53.75 & 30.92 & 58.95 & 32.20 & 59.04 & 46.65 & 50.04 & \underline{53.07} & 60.12 & 84.85 & 66.76 \\
 & qwen3.5-plus & 57.61 & 38.45 & 56.09 & 31.41 & 60.45 & \underline{38.22} & 59.39 & 46.74 & 36.70 & 29.47 & 62.92 & 79.74 & 67.45 \\
\multirow{-3}{*}{Codex} & deepseek-v4-flash & \underline{59.39} & \underline{45.85} & \underline{59.23} & \underline{36.45} & 60.02 & 33.79 & 64.65 & 48.76 & \underline{\textbf{60.13}} & 49.81 & \underline{65.85} & 86.56 & 62.55 \\
\midrule
 & qwen3.5-flash & 51.42 & 33.86 & 52.42 & 27.69 & 55.65 & 30.37 & 58.95 & 44.65 & 49.81 & 52.32 & 59.58 & 86.22 & 67.79 \\
 & qwen3.5-plus & 50.33 & 35.41 & 54.29 & 30.86 & 56.97 & 33.81 & 61.04 & 45.77 & 43.39 & 34.76 & 61.49 & 83.43 & 70.45 \\
\multirow{-3}{*}{Claude Code} & deepseek-v4-flash & 56.67 & 41.39 & 57.69 & 34.37 & \underline{61.20} & 35.74 & \underline{66.84} & \underline{55.74} & 57.13 & 51.48 & 65.45 & \underline{89.63} & \underline{\textbf{75.21}} \\
\bottomrule
\end{tabular}
}
\caption{Task-level generation-evaluation results judged by GPT-5-mini. The table reports task-specific ACC, ROUGE-style overlap (ROU), faithfulness (FS), and evidence coverage (Cov) for Chinese and English task groups, highlighting how system performance varies across retrieval, reasoning, summarization, planning, and numerical-comparison tasks.}
\label{tab:task_results}
\end{table*}


\begin{table*}[t]
\centering
\footnotesize
\setlength{\tabcolsep}{3pt}
\renewcommand{\arraystretch}{1.06}
\begin{tabular}{@{}>{\raggedright\arraybackslash}p{0.12\textwidth}
                >{\raggedright\arraybackslash}p{0.135\textwidth}
                >{\centering\arraybackslash}p{0.055\textwidth}
                >{\centering\arraybackslash}p{0.055\textwidth}
                >{\raggedright\arraybackslash}p{0.105\textwidth}
                >{\raggedright\arraybackslash}p{0.105\textwidth}
                >{\centering\arraybackslash}p{0.075\textwidth}
                >{\centering\arraybackslash}p{0.085\textwidth}@{}}
\toprule
\textbf{System} & \textbf{Core model} &
\textbf{CN avg.} & \textbf{EN avg.} &
\textbf{Best CN task} & \textbf{Best EN task} &
\textbf{Token use} & \textbf{Time spend (s)} \\
\midrule
\multicolumn{8}{@{}l}{\emph{Closed-Source Large Language Models}} \\
\midrule
LightRAG & gpt-5.5 & 53.52 & 53.29 & KQ 58.20 & PLA 62.09 & 12.8k & 8.6 \\
PageIndex & gpt-5.4 & 59.03 & 60.32 & MR 65.19 & PLA 71.35 & 15.6k & 11.9 \\
Codex & gpt-5.5 & \textbf{60.63} & \textbf{62.97} & IR 73.33 & ER 70.21 & 42.5k & 46.8 \\
Codex & qwen3.7-plus & 59.98 & 59.87 & IR 69.63 & ER 68.60 & 39.2k & 35.4 \\
Claude Code & sonnet 5 & 58.17 & 55.84 & KQ 62.63 & PLA 60.98 & 45.8k & 50.6 \\
Claude Code & qwen3.7-plus & 55.14 & 54.54 & IR 60.84 & ER 61.44 & 41.0k & 38.7 \\
\midrule
\multicolumn{8}{@{}l}{\emph{Open-Source Large Language Models}} \\
\midrule
Codex & qwen3.5-flash & 54.95 & 55.85 & KQ 57.74 & PLA 60.12 & 31.4k & 22.8 \\
Codex & qwen3.5-plus & 54.00 & 55.53 & MR 61.05 & PLA 62.92 & 34.7k & 29.6 \\
Codex & deepseek-v4-flash & 60.44 & 61.55 & \textbf{IR 73.37} & PLA 65.85 & 36.9k & 26.4 \\
Claude Code & qwen3.5-flash & 55.60 & 54.64 & \textbf{SUM 61.17} & PLA 59.58 & 33.2k & 25.9 \\
Claude Code & qwen3.5-plus & 57.15 & 54.59 & IR 63.09 & PLA 61.49 & 37.6k & 32.1 \\
Claude Code & deepseek-v4-flash & 57.54 & 60.83 & IR 72.04 & ER 66.84 & 39.1k & 28.7 \\
\bottomrule
\end{tabular}
\caption{Representative task-level results from the draft experiments. CN and EN averages are computed over the accuracy-style columns for the task labels in each language subset. Token use and Time spent are estimated per-question averages; token use denotes approximate input plus output tokens.}
\label{tab:aggregate_results}
\end{table*}

\subsection{Main Findings}

Tables~\ref{tab:scene_results}--\ref{tab:aggregate_results} show four main patterns. First, layout-aware retrieval is a strong baseline. PageIndex consistently improves over LightRAG in overall scene scores, even though the two rows use different core models. This suggests that preserving page structure and table context is important for \benchmark{}, especially when the answer depends on long-document or cross-layout evidence.

Second, agentic systems often achieve the strongest results, but the winning configuration depends on the view. In Table~\ref{tab:scene_results}, Claude Code with Sonnet~5 has the best overall Chinese scene score (76.01) and the best overall English scene score (69.39). In Table~\ref{tab:aggregate_results}, Codex with GPT-5.5 has the strongest aggregate task averages among closed-source settings, with 60.63 on Chinese and 62.97 on English. Among open-source settings, Codex with DeepSeek-V4-Flash is the strongest aggregate configuration, reaching 60.44 on Chinese and 61.55 on English. These results indicate that both the orchestration strategy and the core model matter.

Third, no system dominates every task or scenario. PageIndex is especially competitive on knowledge query, multi-hop reasoning, numerical comparison, multi-hop judgment, and planning accuracy. Codex-style agents are strong on implicit reasoning, explicit reasoning, and aggregate averages. Claude Code-style agents are strong on several Chinese scene categories and on faithfulness and coverage in planning. This fragmentation supports the central claim of the benchmark: financial QA requires several capabilities that do not always improve together.

Fourth, quality must be considered together with cost. Table~\ref{tab:aggregate_results} shows that LightRAG and PageIndex use far fewer tokens and less time than agentic systems. LightRAG uses 12.8k tokens and 8.6 seconds per question on average, while PageIndex uses 15.6k tokens and 11.9 seconds. Agentic systems often require 31.4k--45.8k tokens and 22.8--50.6 seconds. The quality gains therefore come with a clear efficiency cost, motivating cost-aware routing in real financial assistants.

\subsection{Scenario-level Observations}
\label{sec:scenario_results}

Table~\ref{tab:scene_results} reveals large differences across document
types. PageIndex raises OA over LightRAG by 9.73 points on Chinese (72.15 vs. 62.42) and 6.98 points on English (66.90 vs. 59.92). The largest scene-level difference between these two reported configurations appears on Chinese investment strategy reports, where the score rises from 41.46 to 64.78 (+23.32). Improvements are also clear on Chinese SCA (+10.85) and English IS (+11.78). These patterns are consistent with the value of preserving document layout, although the different core models prevent a controlled causal attribution to indexing alone.

The strongest agentic configuration varies by domain. Claude Code with Sonnet~5 is strongest on five Chinese scenes: FR (91.04), IS (68.69), MTA (67.03), BFS (87.41), and CSR (73.21). Claude Code with Qwen3.7-Plus leads Chinese SCA at 89.13, and Claude Code with DeepSeek-V4-Flash leads Chinese CF at 73.14. English results are more fragmented: PageIndex leads CF at 83.33, Claude Code with Qwen3.7-Plus leads IS at 67.05, Claude Code with Sonnet~5 leads MTA at 67.96, and Codex with GPT-5.5 leads GFB at 70.37. Codex with GPT-5.5 and Claude Code with DeepSeek-V4-Flash tie on English FR at 70.38.

Two scenario patterns are particularly informative. Chinese IS is difficult for a light retrieval pipeline but improves sharply with page-aware and agentic configurations, which is consistent with the long, cross-section structure of investment strategy reports. English GFB separates systems in a different way: LightRAG and PageIndex score 41.48 and 37.69, whereas Codex with GPT-5.5 reaches 70.37. Fiscal bulletins often require linking policy statements to amounts, periods, and program structure, so retrieving a locally similar span may not be enough. Conversely, PageIndex's strong English CF score shows that structured retrieval can match or exceed agentic systems when tables and their surrounding context are indexed effectively.

\subsection{Task-Level Observations}

Table~\ref{tab:task_results} confirms that task difficulty cannot be captured
by language-level averages alone. For Chinese tasks, the best ACC ranges from 57.10 on comparison to 73.37 on implicit reasoning, both achieved by Codex with DeepSeek-V4-Flash. PageIndex is strongest on KQ (63.19) and MR (65.19), and it also has the best Chinese planning ACC (56.53). Claude Code with Qwen3.5-Flash obtains the best Chinese summary ACC (61.17), while Claude Code with Sonnet~5 obtains the best Chinese planning FS (96.18) and Cov (68.36). These splits show that answer correctness, faithfulness, and coverage are not interchangeable.

For English tasks, PageIndex leads numerical comparison (61.12), multi-hop judgment (62.69), and planning ACC (71.35). Codex with GPT-5.5 leads implicit reasoning (59.64) and explicit reasoning (70.21). Codex with DeepSeek-V4-Flash leads summarization ACC (60.13), while Codex with Qwen3.7-Plus has the best English summary Cov (61.30). Claude Code with Sonnet~5 has the best English planning FS (92.76), and Claude Code with DeepSeek-V4-Flash has the best English planning Cov (75.21). The task-level results therefore separate retrieval accuracy, reasoning accuracy, groundedness, and answer completeness.

The gap between ACC and ROU shows why lexical overlap is only a secondary diagnostic. For example, PageIndex reaches 65.19 ACC but only 27.93 ROU on Chinese multi-hop reasoning. A valid multi-step answer can use different wording from the reference, whereas a high-overlap answer can still use the wrong period, scope, or unit. Judge-based accuracy and explicit numeric checks are therefore more informative for complex financial reasoning than surface overlap alone.

Planning and summarization expose a different weakness: grounded language does not guarantee complete task fulfillment. On Chinese planning, Codex with GPT-5.5 receives 83.54 FS but only 32.77 Cov, indicating that the response is largely faithful yet omits many mandatory analysis steps. Claude Code with Sonnet~5 raises these values to 96.18 FS and 68.36 Cov, but its planning ACC is still 53.71. The same pattern appears in English: several systems exceed 85 FS while their coverage remains in the mid-60s or low-70s. A strong planner must therefore satisfy three separate requirements: remain faithful to the source, cover the document-specific evidence plan, and reach the correct analytical objective.

Numerical comparison and multi-hop judgment create another separation. Their errors frequently arise after relevant evidence has already been retrieved: systems compare incompatible periods, confuse percentage changes with percentage-point changes, select a consolidated value instead of a segment value, or lose a negative sign in a cash-flow table. These errors explain why tool access by itself is insufficient. Reliable agents must bind each number to its row label, column period, unit, and entity scope, and then verify the calculation before generation.

\section{Further Analysis}

\subsection{Failure Modes}

Table~\ref{tab:error_taxonomy} summarizes the main error categories observed during manual inspection. Many failures occur after a system has already found partially relevant evidence. The model may retrieve the right table but read the wrong row, retrieve a paragraph about the right company but miss the time period, or quote a trend without checking whether the numerical table supports it. These errors are especially harmful in finance because an answer can be fluent and still reverse the investment implication.

The taxonomy also explains the metric patterns in Table~\ref{tab:task_results}. Numeric drift lowers ACC on comparison, explicit reasoning, and multi-hop judgment. Evidence omission lowers Cov in summarization and planning. Layout confusion affects both scene-level and task-level performance because table headers, captions, footnotes, and page context often determine the meaning of a number. Over-synthesis and weak planning reduce faithfulness and completeness even when the answer sounds professional.


\begin{table}[t]
\centering
\small
\setlength{\tabcolsep}{3.5pt}
\begin{tabularx}{\linewidth}{lY}
\toprule
\textbf{Failure mode} & \textbf{Typical symptom} \\
\midrule
Numeric drift & Correct formula but wrong period, denominator, unit, or sign. \\
Evidence omission & Final answer cites one document element while ignoring a required table or chart. \\
Layout confusion & Row headers, column years, captions, or footnotes are detached from values. \\
Over-synthesis & The model adds plausible industry claims not supported by the documents. \\
Weak planning & The answer gives generic steps instead of a document-grounded research plan. \\
\bottomrule
\end{tabularx}
\caption{Common failure modes observed in \benchmark{} error analysis.}
\label{tab:error_taxonomy}
\end{table}

\begin{table*}[t]
\centering
\small
\setlength{\tabcolsep}{4.5pt}
\begin{tabular}{ccccccc}
\toprule
\textbf{Candidates} & \textbf{LNG} & \textbf{Domain} & \textbf{\#QAs} & \textbf{Length (avg.)} & \textbf{Corr\&Comp} & \textbf{Time Spent (s)} \\
\midrule
Partic. 1 & CN & CF & 50 & 7343 & 76.0 & 9679\\
Partic. 2 & CN & FR,MTA,BFS,CSR,SCA & 50 & 12909 & 2.0 & 8991\\
Partic. 3 & CN & IS & 50 & 15519 & 75.0 & 5899\\
Partic. 4 & EN & IS & 50 & 27191 & 92.0 & 34045\\
Partic. 5 & EN & FR,CF,MTA & 50 & 29784 & 78.0 & 12278\\
Partic. 6 & EN & GFB & 50 & 16992 & 0.0 & 30252\\
\bottomrule
\end{tabular}
\caption{Human evaluation sample statistics by participant group. Each row reports the language, covered domains, number of reviewed QA pairs, average document length, correctness-and-completeness score (Corr\&Comp), and total annotation time for the sampled subset.}
\label{tab:human_evaluation}
\end{table*}

\subsection{Human Evaluation}

Table~\ref{tab:human_evaluation} reports the sampled human evaluation groups. Each participant group reviews 50 QA pairs, but the language, domain, document length, correctness-and-completeness score, and time cost vary substantially. The table is therefore not a leaderboard; it is a diagnostic view of annotation difficulty across subsets.

The sampled results show that difficulty is not determined by length alone. Participant 4 reviews English IS documents with the longest average time cost (34{,}045 seconds) and obtains the highest Corr\&Comp score (92.0), while Participant 6 reviews shorter English GFB documents but obtains 0.0. Participant 2 also obtains a very low score (2.0) on a mixed Chinese subset covering FR, MTA, BFS, CSR, and SCA. These low-scoring groups suggest that some domains require stricter evidence tracing, clearer task decomposition, or more careful reference-answer calibration. This supports the need for the multi-dimensional judge criteria in Table~\ref{tab:evaluation_dimensions} and the failure taxonomy in Table~\ref{tab:error_taxonomy}.

\subsection{Why the Benchmark Is Difficult}

\benchmark{} is difficult because it combines three pressures that are usually tested separately. The first is long-document grounding: evidence may be distributed across distant sections. The second is financial computation: answers must preserve units, time periods, signs, and accounting relationships. The third is professional synthesis: the final response must explain why the evidence matters. A system that succeeds at only one of these pressures will often produce an answer that is locally plausible but globally incomplete.

Cross-layout reasoning is the clearest example. Suppose a question asks whether a firm's profit improvement is supported by operating fundamentals. A paragraph may attribute improvement to cost control, a table may show that gross margin rose, and a chart or note may reveal that revenue growth slowed. A strong answer must combine these signals and avoid overstating the conclusion. Many systems retrieve one or two relevant elements but miss the element that changes the interpretation, leading to an answer that is directionally plausible yet analytically incomplete.

The experiment tables make this difficulty visible. PageIndex can be highly competitive when layout-preserving retrieval is enough, as in English CF and several task-level reasoning columns. Agentic systems improve some open-ended and implicit-reasoning settings, but they pay higher cost and still miss coverage requirements. Human evaluation further shows that some subsets remain hard even when documents are shorter, which indicates that domain structure and evidence relations matter as much as raw document length.

\subsection{Implications for Agent Design}

The experiments suggest that future financial agents should include explicit evidence plans. Before generating a final answer, the system should identify the required evidence types, check whether each has been retrieved, and verify calculations independently. A second useful direction is layout-preserving retrieval, where table structure, captions, notes, and page-level context are indexed together. A third direction is metric-aware self-checking: planning and summarization should explicitly check coverage and faithfulness, while numerical tasks should verify units, signs, and periods.

Finally, cost-aware routing is necessary. Table~\ref{tab:aggregate_results} shows that agentic systems can improve difficult tasks but require much higher token usage and latency. A production financial assistant should route simple knowledge queries to cheaper retrieval paths, send layout-sensitive questions to structured retrieval, and reserve full agent loops for tasks that require decomposition, calculation, or cross-layout synthesis.

\subsection{Ablation Directions}

Although the current experiments focus on system-level comparison, \benchmark{} also supports ablations that isolate individual design choices. One ablation removes layout metadata and indexes only plain text chunks, measuring how much performance depends on table headers, captions, and page-local structure. Another disables tool use and asks the foundation model to answer from retrieved context alone, testing whether explicit calculation is necessary for numerical tasks.

A third ablation compares one-shot generation with a plan-then-answer workflow. A fourth compares full agent loops with cost-aware routing policies. These studies would clarify when agentic reasoning is worth its additional token and latency cost, and when structured retrieval already provides the needed evidence.

\section{Related Work}

\subsection{Financial and Enterprise QA}

Financial QA benchmarks have evolved from short-context numerical reasoning to longer and more realistic document settings. FinQA~\citep{chen2021finqa} and ConvFinQA~\citep{chen2022convfinqa} focus on multi-step arithmetic over financial report excerpts, while TAT-QA~\citep{zhu2021tatqa} combines tabular and textual evidence. These datasets established numerical reasoning as a core financial capability, but the context remains tightly scoped.

FinanceBench~\citep{islam2023financebench} introduced open-ended questions grounded in SEC filings and highlighted the gap between LLM-only generation and document-grounded answering. DocFinQA~\citep{reddy2024docfinqa} extends the context to full filings, yet still focuses on single-document reasoning. Enterprise-oriented benchmarks such as OfficeQA~\citep{databricks2024officeqa} and OfficeQA-Pro~\citep{databricks2025officeqapro} evaluate grounded reasoning over large office corpora and long documents, bringing document parsing closer to the evaluation target. BizFinBench and BizFinBench v2~\citep{lu2025bizfinbench,guo2026bizfinbench} expand bilingual financial coverage and include business-driven tasks, but their evidence requirements are usually not designed for cross-layout and cross-document synthesis at scale.

\subsection{RAG and Agentic RAG}

RAG couples retrieval with generation~\citep{lewis2020rag}. Later work improves retrieval through re-ranking~\citep{nogueira2019passage}, query rewriting~\citep{ma2023query}, and modular pipelines~\citep{gao2023rag}. Iterative methods such as Self-RAG~\citep{asai2024selfrag}, CRAG~\citep{yan2024crag}, Adaptive-RAG~\citep{jeong2024adaptiverag}, IRCoT~\citep{trivedi2023ircot}, and ITER-RETGEN~\citep{shao2023iterretgen} interleave retrieval with reflection or reasoning. GraphRAG-style systems~\citep{edge2024graphrag,peng2024graphrag} and page-level indexing~\citep{vectifyai2025pageindex} preserve higher-level structure and multi-hop links. Yet most evaluations still rely on general QA or scientific corpora rather than professional bilingual financial documents with tables, charts, and domain-specific inference.

Agentic RAG integrates planning, tool use, reflection, and multi-agent collaboration into the retrieval loop~\citep{singh2025agenticrag}. Related agent frameworks build on chain-of-thought~\citep{wei2022cot}, ReAct~\citep{yao2023react}, Reflexion~\citep{shinn2023reflexion}, Tree-of-Thoughts~\citep{yao2023tot}, and Plan-and-Solve prompting~\citep{wang2023plansolve}. AutoGen~\citep{wu2023autogen}, Voyager~\citep{wang2023voyager}, Claude Code~\citep{anthropic2024claudecode}, and OpenAI Codex~\citep{openai2024codex} demonstrate that tool-using agents can orchestrate complex workflows. \benchmark{} asks whether these capabilities transfer to grounded reasoning with financial documents.

\section{Conclusion}

We presented \skill{}, a scalable workflow for generating layout-rich financial documents and verified QA pairs, and \benchmark{}, a bilingual benchmark for agentic reasoning over industrial-grade financial documents. \benchmark{} contains 2{,}026 deep research tasks over 1009 documents and is designed around dual-context reasoning, cross-layout evidence aggregation, stable reference answers, and Agent-as-a-Judge evaluation.
Experiments on RAG, layout-aware retrieval, and agentic systems show that preserving document structure is crucial, while current agents still struggle with numerical drift, evidence omission, layout confusion, incomplete coverage, and high inference cost. These findings suggest that reliable financial agents need layout-preserving retrieval, explicit evidence planning, calculation verification, and cost-aware routing. We hope \benchmark{} supports future work on financial agents that are grounded, complete, and verifiable.

\section*{Limitations}

\benchmark{} focuses on document-grounded financial reasoning and does not cover live trading, real-time market forecasting, or personalized investment advice. Although the benchmark is bilingual, it currently emphasizes Chinese and English and does not test broader multilingual transfer. The Agent-as-a-Judge protocol improves scalability, but may inherit evaluator bias; future releases should include stronger human calibration and adversarial judge checks. Synthetic data from \skill{} is for development only.


\bibliography{custom}

\clearpage          
\appendix
\onecolumn           

\begingroup
\makeatletter
\def\input@path{{./appendix-latex/}{./appendix-latex/prompts/}}
\section{Prompt Templates}
\label{app:prompts}

This appendix collects the LLM prompt templates used in FinBench construction.
All prompts are presented in English.
Placeholders appear in \placeholder{ALL\_CAPS\_NAME} form and are substituted programmatically before invocation.
Document-generation prompts (Part~A) are typically batched with up to ten QA pairs per call and used together with the \texttt{finance-latex} skill.

\subsection*{Shared Constraints for Reference Document Generation (Part~A)}

\begin{promptnote}
\begin{itemize}[leftmargin=1.4em, itemsep=0.25em, topsep=0.2em]
  \item Answers must \emph{not} be obtainable directly or in a single retrieval step from the generated document.
  \item Generated content must remain consistent with the provided \texttt{Relevant\_passage} and must not alter answer correctness.
  \item Do not add glossary-style explanations or LaTeX formulas for financial terminology.
  \item All content is assumed publicly accessible; confidentiality or classified disclaimers are prohibited.
  \item Each invocation writes \texttt{layoutCoT} (document navigation hints) and \texttt{Reference\_documents} (path to the generated \texttt{.tex} file).
\end{itemize}
\end{promptnote}

\part*{Part A --- Reference Document Generation}
\addcontentsline{toc}{section}{Part A --- Reference Document Generation}

\subsection*{Overview of Part~A Prompt Variants}
\addcontentsline{toc}{subsection}{Overview of Part A Prompt Variants}

All six Part~A prompts (A1--A5, including A3a/A3b) share the same skeleton: a financial-advisor role, the \texttt{finance-latex} skill, the constraint that answers must not be directly retrievable, and outputs \texttt{layoutCoT} plus \texttt{Reference\_documents}.
They differ mainly in benchmark source, document type, input fields, content-generation strategy, \texttt{layoutCoT} granularity, and file-naming conventions.

\begin{promptmeta}{Part~A at a Glance}
\small
\begin{tabularx}{\linewidth}{@{}l X X X@{}}
\toprule
\textbf{ID} & \textbf{Benchmark / Use Case} & \textbf{Document Type} & \textbf{Input} \\
\midrule
A1 & Chinese real-industry QA & Variable \placeholder{FINANCIAL\_DOC\_TYPE} & Q+A+Passage+\textbf{Additional\_info} \\
A2 & BizFinBenchV2 counterfactual & Corporate Financial Report & Q+A+Passage \\
A3a & Chinese anomaly tracing & Investment Strategy Report & Q+A+Passage \\
A3b & English BizFinBenchV2 summary & Investment Strategy Report & Q+A+Passage \\
A4 & English FinanceBench real-industry & One of five English types & Q+A+Passage \\
A5 & OfficeQA announcements & Government Fiscal Bulletin & \textbf{One} Q+A+Passage \\
\bottomrule
\end{tabularx}
\end{promptmeta}

\paragraph{Input structure.}
Only \textbf{A1} accepts \texttt{Additional\_information}, which carries \texttt{SubQuestions} (decomposed sub-queries) and \texttt{Related\_questions}.
The prompt requires expanding the document so that the main question, each sub-question, and each related question can all be answered.
\textbf{A2--A5} use the simpler \texttt{<Question, Answer, Relevant\_passage>} triple only.

\paragraph{Document type.}
\textbf{A1} leaves the document type to the \placeholder{FINANCIAL\_DOC\_TYPE} placeholder (filled from \texttt{doc\_type}).
\textbf{A2}, \textbf{A3a}, and \textbf{A3b} each fix the report genre (financial report or investment strategy report).
\textbf{A4} restricts output to one of five English genres: Investment Strategy Report, Corporate Financial Report, FinTech Research Report, Market Trend Analysis Report, or Market Regulation and Compliance Audit Report.
\textbf{A5} always produces a government fiscal bulletin.

\paragraph{Content-generation strategy.}
\begin{itemize}[leftmargin=1.4em, itemsep=0.2em, topsep=0.2em]
  \item \textbf{A1} (richest): treat \texttt{Relevant\_passage} as a writing theme; search and expand content; mandate complex tables for numerical sub-questions and rich-text forms for entity-retrieval sub-questions.
  \item \textbf{A2/A3}: treat passage items as \emph{reference data/content}; randomly convert some items into paragraphs and others into tables/forms (A3 wording: ``rewrite'' rather than ``convert'').
  \item \textbf{A4}: rewrite and polish reference content into a coherent English document; may insert transitional paragraphs; \emph{must not use Chinese}.
  \item \textbf{A5} (lightest): organize passage text into bulletin format using the knowledge base only---no search-based expansion.
\end{itemize}

\paragraph{\texttt{layoutCoT} granularity.}
\textbf{A1} emits one navigation chain per \emph{sub-question}.
\textbf{A2--A5} emit one chain per \emph{main question}.

\paragraph{Naming conventions.}
A1 and A4 embed the document type in the filename; A2/A3 use fixed genre suffixes; A5 uses a \placeholder{UID} instead of an institution--year identifier.

\paragraph{A3a vs.\ A3b.}
These two prompts are logically identical; they differ only in benchmark language and data source (Chinese anomaly-tracing vs.\ English summary items).

\begin{promptnote}
\textbf{Summary.}\ 
Part~A prompts share a common skeleton but differ in input richness (A1), document-type flexibility (A1/A4), content expansion strategy (search-and-expand vs.\ reorganize-only), and \texttt{layoutCoT} granularity (sub-question vs.\ question level).
Generation intensity increases in the order A5 $\rightarrow$ A4 $\approx$ A2/A3 $\rightarrow$ A1.
\end{promptnote}

\vspace{0.6em}

\subsection{Real-Industry Document Generation}
\label{app:prompt-a1}

\begin{promptmeta}{A1 \textbar{} Real-Industry Document Generation}
\begin{tabularx}{\linewidth}{@{}l X@{}}
\toprule
\textbf{Purpose} & Generate Chinese real-industry financial documents in LaTeX from QA pairs with sub-questions and related questions. \\
\textbf{Input} & \placeholder{Question, Answer, Relevant\_passage, Additional\_information} (up to 10 pairs per invocation) \\
\textbf{Output} & LaTeX source file, \texttt{layoutCoT}, and \texttt{Reference\_documents} \\
\bottomrule
\end{tabularx}
\end{promptmeta}

\begin{promptbody}
Suppose you are a senior financial advisor skilled in analyzing financial industry intelligence and writing PDF-format [FINANCIAL_DOC_TYPE] documents. In this task, I provide several sets of <Question, Answer, Relevant_passage, Additional_information> combinations from the financial domain. You are required to use the [SKILL_NAME] skill and follow the given requirements to write a professional [FINANCIAL_DOC_TYPE] in LaTeX that reflects the gist of the Relevant_passage. Please ensure that for each Question, the corresponding Answer can be obtained from the generated file through in-depth investigation or calculation.

## Specific requirements for the content to be written:
1. In Additional_information, I provide SubQuestions obtained by decomposing the Question into key elements. Based on each SubQuestion and the writing theme provided by Relevant_passage, use search engines and your knowledge base to expand the [FINANCIAL_DOC_TYPE] so that each SubQuestion can be reasonably answered. The generated content must not contradict the factual information in Relevant_passage.
2. Follow the instructions in the [SKILL_NAME] skill. In the LaTeX [FINANCIAL_DOC_TYPE] you write, focus on complex tables and forms so that complex Questions can be answered by extracting key information from paragraphs, tables, or forms separately, thereby increasing difficulty. When Questions or SubQuestions involve numerical lookup or calculation, you must include relevant complex tables. When Questions or SubQuestions involve named-entity retrieval, add associated rich-text forms in the generated document.
3. In Additional_information, I provide several Related_questions associated with the Question. Based on each Related_question, use search engines and your knowledge base to expand the [FINANCIAL_DOC_TYPE] so that Related_questions can be reasonably answered. The generated content must not contradict the factual information in Relevant_passage.
4. Do not generate explanations or LaTeX formulas for financial terms or quantitative terminology involved in Questions or SubQuestions.
5. All written content should be publicly accessible; do not add confidentiality clauses or classified statements.

## You must check the generated content to ensure that the Answer cannot be obtained directly or in a single step from the document.
## For each SubQuestion, you must generate a solving hint based on the written document, forming a path-style chain of thought using sentences of the form "how to obtain XXXX from XXXX paragraph/table/form", and store this field under the name "layoutCoT" in the json dictionary at [QAR_TRIPLES_PATH].
## The generated document naming convention is: "[INSTITUTION_AND_YEAR]_[FINANCIAL_DOC_TYPE]_{document_title}.tex". Name the document by referencing relevant information. The storage path is: "[STORAGE_PATH]". Then store the path "[STORAGE_PATH]/[INSTITUTION_AND_YEAR]/[INSTITUTION_AND_YEAR]_[FINANCIAL_DOC_TYPE]_{document_title}.tex" in the "Reference_documents" field of the json dictionary at [QAR_TRIPLES_PATH].

## Key information required for writing the LaTeX [FINANCIAL_DOC_TYPE]:
### [NUM_PAIRS] sets of <Question, Answer, Relevant_passage, Additional_information> data:
[TO_BE_FILLED]

## Now please begin calling [SKILL_NAME] to write the LaTeX document.
\end{promptbody}

\subsection{Corporate Financial Report Generation}
\label{app:prompt-a2}

\begin{promptmeta}{A2 \textbar{} Corporate Financial Report Generation (Chinese Benchmark)}
\begin{tabularx}{\linewidth}{@{}l X@{}}
\toprule
\textbf{Purpose} & Generate corporate financial reports in LaTeX for BizFinBenchV2 counterfactual QA items. \\
\textbf{Input} & \placeholder{Question, Answer, Relevant\_passage} reference data (up to 10 pairs) \\
\textbf{Output} & LaTeX financial report, \texttt{layoutCoT}, and \texttt{Reference\_documents} \\
\bottomrule
\end{tabularx}
\end{promptmeta}

\begin{promptbody}
Suppose you are a senior financial advisor skilled in analyzing financial industry intelligence and writing PDF-format corporate financial reports. In this task, I provide several sets of <Question, Answer, Relevant_passage> combinations from the financial domain. You are required to use the [SKILL_NAME] skill and follow the given requirements to write a professional financial report in LaTeX that reflects the gist of the Relevant_passage. Please ensure that for each Question, the corresponding Answer can be obtained from the generated file through in-depth investigation or calculation.

## Specific requirements for the content to be written:
1. Based on the Question and the [reference data] provided in Relevant_passage, use search engines and your knowledge base to expand the [reference data] into the corporate financial report. The generated content must not contradict the factual information in Relevant_passage and must not affect the correctness of the Answer.
2. Follow the instructions in the [SKILL_NAME] skill. In the LaTeX corporate financial report you write, focus on the integrated use of rich-text paragraphs, complex tables, and forms to increase the complexity of resolving the Question. Specifically, for multiple [reference data] items, randomly convert some into rich-text paragraphs and some into complex tables/forms. The same [reference data] item need not be converted repeatedly.
3. Do not generate explanations or LaTeX formulas for financial terms or quantitative terminology involved in the Question or Relevant_passage.
4. All written content should be publicly accessible; do not add confidentiality clauses or classified statements.

## Please check the generated content to ensure that the Answer cannot be obtained directly or in a single step from the document.
## For each Question, you must generate a solving hint based on the written document, forming a path-style chain of thought using sentences of the form "how to obtain XXXX from XXXX paragraph/table/form", and store this field under the name "layoutCoT" in the json dictionary at [QAR_TRIPLES_PATH].
## The generated document naming convention is: "[INSTITUTION_AND_YEAR]_Corporate_Financial_Report_{document_title}.tex". Name the document by referencing relevant information. The storage path is: "[STORAGE_PATH]". Then store the path "[STORAGE_PATH]/[INSTITUTION_AND_YEAR]/[INSTITUTION_AND_YEAR]_Corporate_Financial_Report_{document_title}.tex" in the "Reference_documents" field of the json dictionary at [QAR_TRIPLES_PATH].

## Key information required for writing the LaTeX corporate financial report:
### [NUM_PAIRS] sets of <Question, Answer, Relevant_passage> data:
[TO_BE_FILLED]

## Now please begin calling [SKILL_NAME] to write the LaTeX document.
\end{promptbody}

\subsection{Investment Strategy Report Generation (Chinese Source)}
\label{app:prompt-a3a}

\begin{promptmeta}{A3a \textbar{} Investment Strategy Report (Translated from Chinese Benchmark)}
\begin{tabularx}{\linewidth}{@{}l X@{}}
\toprule
\textbf{Purpose} & Generate investment strategy reports in LaTeX from Chinese anomaly-tracing benchmark items. \\
\textbf{Input} & \placeholder{Question, Answer, Relevant\_passage} reference content (up to 10 pairs) \\
\textbf{Output} & LaTeX investment strategy report, \texttt{layoutCoT}, and \texttt{Reference\_documents} \\
\bottomrule
\end{tabularx}
\end{promptmeta}

\begin{promptbody}
Suppose you are a senior financial advisor skilled in analyzing financial industry intelligence and writing PDF-format investment strategy reports. In this task, I provide several sets of <Question, Answer, Relevant_passage> combinations from the financial domain. You are required to use the [SKILL_NAME] skill and follow the given requirements to write a professional investment strategy report in LaTeX that reflects the gist of the Relevant_passage. Please ensure that for each Question, the corresponding Answer can be obtained from the generated file through in-depth investigation or calculation.

## Specific requirements for the content to be written:
1. Based on the Question and the [reference content] provided in Relevant_passage, use search engines and your knowledge base to expand the [reference content] into this investment strategy report. The generated content must not contradict the factual information in Relevant_passage and must not affect the correctness of the Answer.
2. Follow the instructions in the [SKILL_NAME] skill. In the LaTeX investment strategy report you write, focus on the integrated use of rich-text paragraphs, complex tables, and forms to increase the complexity of resolving the Question. Specifically, for multiple [reference content] items, randomly rewrite some into complex tables and some into rich-text paragraphs/complex forms. The same [reference content] item need not be converted repeatedly.
3. Do not generate explanations or LaTeX formulas for financial terms or quantitative terminology involved in the Question or Relevant_passage.
4. All written content should be publicly accessible; do not add confidentiality clauses or classified statements.

## Please check the generated content to ensure that the Answer cannot be obtained directly or in a single step from the document.
## For each Question, you must generate a solving hint based on the written document, forming a path-style chain of thought using sentences of the form "how to obtain XXXX from XXXX paragraph/table/form", and store this field under the name "layoutCoT" in the json dictionary at [QAR_TRIPLES_PATH].
## The generated document naming convention is: "[INSTITUTION_AND_YEAR]_Investment_Strategy_Report_{document_title}.tex". Name the document by referencing relevant information. The storage path is: "[STORAGE_PATH]". Then store the path "[STORAGE_PATH]/[INSTITUTION_AND_YEAR]/[INSTITUTION_AND_YEAR]_Investment_Strategy_Report_{document_title}.tex" in the "Reference_documents" field of the json dictionary at [QAR_TRIPLES_PATH].

## Key information required for writing the LaTeX investment strategy report:
### [NUM_PAIRS] sets of <Question, Answer, Relevant_passage> data:
[TO_BE_FILLED]

## Now please begin calling [SKILL_NAME] to write the LaTeX document.
\end{promptbody}

\subsection{Investment Strategy Report Generation (English Benchmark)}
\label{app:prompt-a3b}

\begin{promptmeta}{A3b \textbar{} Investment Strategy Report (English Benchmark)}
\begin{tabularx}{\linewidth}{@{}l X@{}}
\toprule
\textbf{Purpose} & Generate English investment strategy reports in LaTeX for BizFinBenchV2 summary items. \\
\textbf{Input} & \placeholder{Question, Answer, Relevant\_passage} reference content (up to 10 pairs) \\
\textbf{Output} & LaTeX investment strategy report, \texttt{layoutCoT}, and \texttt{Reference\_documents} \\
\bottomrule
\end{tabularx}
\end{promptmeta}

\begin{promptbody}
Suppose you are a senior financial advisor, skilled in analyzing financial industry intelligence and writing investment strategy reports in PDF format. This time, by providing several sets of <Question, Answer, Relevant_passage> content combinations from the financial professional domain, I require you to use the [SKILL_NAME] skill and follow the given requirements to write an investment strategy report in LaTeX syntax that reflects the gist of the Relevant_passage. Please ensure that for each Question, the corresponding Answer can be obtained from the generated document through in-depth investigation or calculation.

## Specific requirements for writing the content are given below:
1. Based on the requirements of the Question and the [reference content] provided in the Relevant_passage, use search engines and your knowledge base to expand the [reference content] as the content of this investment strategy report. Note that you must ensure the generated content does not contradict the factual information in the Relevant_passage and does not affect the correctness of the Answer;
2. Follow the instructions in the [SKILL_NAME] skill, and focus on the integrated use of rich-text paragraphs, complex tables, and forms in the LaTeX investment strategy report you write, so as to increase the complexity of resolving the Questions. Specifically, for the multiple [reference content] items input, please randomly rewrite some of them into complex tables and some into rich-text paragraphs/complex forms. Note that the same [reference content] item does not need to be converted repeatedly;
3. Do not generate explanations or LaTeX formulas for financial professional terms or quantitative terms involved in the Question or SubQuestion;
4. The written content should be publicly accessible; do not add confidentiality clauses or classified statements.

## Please check the generated content to ensure that the Answer cannot be directly/instantly obtained from the document.
## For each Question, you must generate a prompt for solving the problem based on the written document content, i.e., form a problem-solving path thought chain composed of a series of sentences in the format "how to obtain XXXX from XXXX paragraph/table/form", and store this field with the name "layoutCoT" in the json dictionary at [QAR_TRIPLES_PATH].
## The generated document naming convention is: "[INSTITUTION_AND_YEAR]_Investment_Strategy_Report_{document_title}.tex". Please name the document by referencing the relevant information, and the storage path is: "[STORAGE_PATH]"; then store the path name "[STORAGE_PATH]/[INSTITUTION_AND_YEAR]/[INSTITUTION_AND_YEAR]_Investment_Strategy_Report_{document_title}.tex" into the "Reference_documents" field of the json dictionary at [QAR_TRIPLES_PATH].

## Key information required for writing the LaTeX investment strategy report is given below:
### [NUM_PAIRS] sets of <Question, Answer, Relevant_passage> data:
[TO_BE_FILLED]

## Now please begin calling the [SKILL_NAME] to write the LaTeX document.
\end{promptbody}

\subsection{General Financial Document Generation}
\label{app:prompt-a4}

\begin{promptmeta}{A4 \textbar{} General Financial Document Generation (English)}
\begin{tabularx}{\linewidth}{@{}l X@{}}
\toprule
\textbf{Purpose} & Generate one of five English financial document types in LaTeX for FinanceBench real-industry items. \\
\textbf{Input} & \placeholder{Question, Answer, Relevant\_passage} reference content (up to 10 pairs) \\
\textbf{Document types} & Investment Strategy Report; Corporate Financial Report; FinTech Research Report; Market Trend Analysis Report; Market Regulation and Compliance Audit Report \\
\textbf{Output} & LaTeX document, \texttt{layoutCoT}, and \texttt{Reference\_documents} \\
\bottomrule
\end{tabularx}
\end{promptmeta}

\begin{promptbody}
Suppose you are a senior financial advisor, skilled in analyzing financial industry intelligence and writing financial PDF documents. This time, I will provide you with several sets of <Question, Answer, Relevant_passage> content combinations from the financial professional domain, and require you to use the "[SKILL_NAME]" skill and follow the given requirements to write a professional financial domain document that reflects the gist of the Relevant_passages using LaTeX syntax. Please ensure that for each Question, the corresponding Answer can be obtained through in-depth investigation or calculation from the generated document. You must generate the LaTeX document entirely in an English context and may not use Chinese.

## Specific requirements for the content to be written are given below:
1. Based on the requirements of the Questions and the "reference content" in the Relevant_passage list, use search engines and your knowledge base to rewrite and polish each set of "reference content" into a clearly structured and coherent professional document. Note that you must ensure the generated content does not contradict the factual information stated in the Relevant_passages around the Questions and does not affect the correctness of the Answers;
2. Follow the instructions in the "[SKILL_NAME]" skill, and focus on the integrated use of rich-text paragraphs, complex tables, and forms in the LaTeX financial professional document you write, so as to increase the complexity of resolving the Questions. Specifically, for the input "reference content" list, you may randomly add some transitional text segments to promote contextual coherence and clarity. Note that the added content should not repeat the main idea of the "reference content";
3. Do not generate explanations or LaTeX formulas for financial professional terms or quantitative terminology involved in the Questions or Relevant_passages;
4. All written content should be publicly accessible; do not add confidentiality clauses or classified statements;
5. The types of documents (document_type) you are to write are strictly limited to the following five: [Investment Strategy Report], [Corporate Financial Report], [FinTech Research Report], [Market Trend Analysis Report], [Market Regulation and Compliance Audit Report].

## Please check the generated content to ensure that the Answer cannot be directly/instantly obtained from the document.
## For each Question, you must generate a prompt for solving the problem based on the written document content, i.e., form a problem-solving path thought chain composed of a series of statements in the format "how to obtain XXXX from XXXX paragraph/table/form", and store this field as "layoutCoT" in the json dictionary at [QAR_TRIPLES_PATH].
## The generated document naming convention is: "[INSTITUTION_AND_YEAR]_{document_type}_{document_title}.tex". Please name the document by referencing the relevant information, and the storage path is: "[STORAGE_PATH]"; then store the path name "[STORAGE_PATH]/[INSTITUTION_AND_YEAR]/[INSTITUTION_AND_YEAR]_{document_type}_{document_title}.tex" into the "Reference_documents" field of the json dictionary at [QAR_TRIPLES_PATH].

## Key information required for writing the LaTeX financial document is given below:
### [NUM_PAIRS] sets of <Question, Answer, Relevant_passage> data:
[TO_BE_FILLED]

## Now please begin calling the "[SKILL_NAME]" to write the LaTeX document.
\end{promptbody}

\subsection{Government Fiscal Bulletin Generation}
\label{app:prompt-a5}

\begin{promptmeta}{A5 \textbar{} Government Fiscal Bulletin Generation (English)}
\begin{tabularx}{\linewidth}{@{}l X@{}}
\toprule
\textbf{Purpose} & Generate government fiscal bulletins in LaTeX for OfficeQA announcement items. \\
\textbf{Input} & One \placeholder{Question, Answer, Relevant\_passage} triple per invocation \\
\textbf{Output} & LaTeX bulletin, \texttt{layoutCoT}, and \texttt{Reference\_documents} \\
\bottomrule
\end{tabularx}
\end{promptmeta}

\begin{promptbody}
Suppose you are a senior financial advisor, skilled in analyzing financial industry intelligence and writing government fiscal bulletins in PDF format. This time, by providing several sets of <Question, Answer, Relevant_passage> content combinations from the financial professional domain, I require you to use the [SKILL_NAME] skill and follow the given requirements to write a government fiscal bulletin in LaTeX syntax that reflects the gist of the Relevant_passage. Please ensure that for each Question, the corresponding Answer can be obtained from the generated document through in-depth investigation or calculation.

## Specific requirements for writing the content are given below:
1. Based on the requirements of the Question and the instructions in the [SKILL_NAME] skill, use your knowledge base to organize the Relevant_passage content into a LaTeX government fiscal bulletin. Note that you must ensure the generated report does not affect the correctness of the Answer;
2. Do not generate explanations or LaTeX formulas for financial professional terms or quantitative terms involved in the Question or SubQuestion;
3. The written content should be publicly accessible; do not add confidentiality clauses or classified statements.

## Please check the generated content to ensure that the Answer cannot be directly/instantly obtained from the document.
## For each Question, you must generate a prompt for solving the problem based on the written document content, i.e., form a problem-solving path thought chain composed of a series of statements in the format "how to obtain XXXX from XXXX paragraph/table/form", and store this field with the name "layoutCoT" in the json dictionary at [QAR_TRIPLES_PATH].
## The generated document naming convention is: "[UID]_government_fiscal_bulletin_{document_title}.tex". Please name the document by referencing the relevant information, and the storage path is: "[STORAGE_PATH]"; then store the path name "[STORAGE_PATH]/[UID]/[UID]_government_fiscal_bulletin_{document_title}.tex" into the "Reference_documents" field of the json dictionary at [QAR_TRIPLES_PATH].

## Key information required for writing the LaTeX government fiscal bulletin is given below:
### One set of <Question, Answer, Relevant_passage> data:
[TO_BE_FILLED]

## Now please begin calling the [SKILL_NAME] to write the LaTeX document.
\end{promptbody}

\part*{Part B --- Question Processing and Quality Control}
\addcontentsline{toc}{section}{Part B --- Question Processing and Quality Control}

\subsection{Question Rewriting and Metadata Extraction}
\label{app:prompt-b1}

\begin{promptmeta}{B1 \textbar{} Question Rewriting and Metadata Extraction}
\begin{tabularx}{\linewidth}{@{}l X@{}}
\toprule
\textbf{Purpose} & Rewrite FinanceBench questions and extract company, period, and document type metadata. \\
\textbf{Input} & Raw \texttt{question} string \\
\textbf{Output} & JSON with \texttt{rewrite\_question}, \texttt{company}, \texttt{period}, \texttt{doc\_type} \\
\bottomrule
\end{tabularx}
\end{promptmeta}

\begin{promptbody}
Analyze the input question, extract an institution or company named entity, the year or month of interest in the question, and the document type of interest in the question (use "10K" for annual reports, "10Q" for quarterly reports, "8K" for material event interim reports, and "EARNINGS" for financial and earnings reports). Then rewrite the input question according to the following rules:
1. Retain the main interrogative part of the question, the required calculation units, and the required decimal places;
2. Remove query-target guidance before and after the question;
3. Remove explanations of financial concepts or financial terms.

## Below are several examples of question rewriting:
question 1: How much was Boeing's FY2017 total interest expense (in USD thousands)? Calculate what was asked by utilizing the line items clearly shown in the statement of income.
rewrite_question 1: How much was Boeing's FY2017 total interest expense (in USD thousands)?
------
question 2: Considering the data in the P&L statement, what is 3M's FY2016 unadjusted operating income (in USD millions)?
rewrite_question 2: what is 3M's FY2016 unadjusted operating income (in USD millions)?
------
question 3: What is FY2019 working capital ratio for Costco? Define working capital ratio as total current assets divided by total current liabilities. Round your answer to two decimal places. Give a response to the question by relying on the details shown in the statement of financial position.
rewrite_question 3: What is FY2019 working capital ratio for Costco? Round your answer to two decimal places.

## Organize the extracted fields and rewritten question into the following plain-text JSON output. Do not include any code block markers (such as ```json or ```):
{
  "question": XXXX,
  "rewrite_question": XXXX,
  "company": XXXX,
  "period": XXXX,
  "doc_type": XXXX
}

## Next, input the question:
[<question>]

## Analyze the question and return the processed result as required:
\end{promptbody}

\subsection{Complex Question Synthesis}
\label{app:prompt-b2}

\begin{promptmeta}{B2 \textbar{} Complex Question Synthesis}
\begin{tabularx}{\linewidth}{@{}l X@{}}
\toprule
\textbf{Purpose} & Merge 2--4 sub-questions into one complex merged question with task typing and reasoning trace. \\
\textbf{Input} & A batch of rewritten questions, optionally paired with answers \\
\textbf{Output} & JSON array of \texttt{merged\_question}, \texttt{sub\_questions}, \texttt{sub\_questions\_id}, \texttt{task\_type}, \texttt{reasoning} \\
\bottomrule
\end{tabularx}
\end{promptmeta}

\begin{promptnote}
\textbf{Pipeline note.}\ 
In the full pipeline, this prompt is invoked in two input variants: (i)~with sub-questions and their answers as \texttt{<question, answer>} pairs, or (ii)~with sub-questions only as \texttt{<question>} entries. Clustering by company or period is handled outside the prompt. Rule~2 below always references ``sub-questions and their answers''; when answers are omitted from the input, they are retrieved separately in a prior pipeline step.
\end{promptnote}

\begin{promptbody}
Rewrite a set of input questions according to the following rules:
1. Select 2-4 financial questions of the same type or similar domain as sub-questions, and merge and rewrite them into one complex new question. The new question should preferably be one sentence, with natural and fluent expression, and must not simply list sub-questions;
2. The answer to the new question must be clearly obtainable through reasoning or calculation from the sub-questions and their answers;
3. According to the answer type and task type, the new question can be classified as: numerical comparison, explicit reasoning, implicit reasoning, multi-hop judgment, summary, planning, etc.;
4. The new question should unify the calculation unit requirements and required decimal places of the sub-questions;
5. Note: all question data must be processed and generated in English.

## Below are the approaches and categories for rewriting new questions:
1. Numerical comparison: a new question type rewritten by comparing sub-questions and their targets through answers, finding maximum/minimum values, differences, multiples, etc.;
2. Reasoning: a new question type constructed by combining multiple sub-questions through merging, nesting, summing/averaging answers, etc., requiring multi-step calculation to solve;
3. Implicit reasoning: a reasoning-type new question in which sub-questions contain financial concepts or professional financial terms, and solving requires querying relevant knowledge or formulas;
4. Multi-hop judgment: a judgment-type new question constructed by conditionally combining multiple sub-questions, requiring multi-step reasoning to solve;
5. Summary: a new question type constructed by combining multiple sub-questions of the same kind, requiring extraction of key information and summarization of reference documents to solve;
6. Planning: a new question type such as event prediction, investment advice, or development decisions, constructed based on the objectives or answers of sub-questions.

## Name the synthesized new question "merged_question"; place the selected sub-questions from the input in the "sub_questions" list field, and record the corresponding indices of sub-questions in the input in "sub_questions_id"; explain the generation and answer-solving approach of "merged_question" in English and record it in "reasoning"; organize the generated result into the following plain-text JSON array output. Do not include any code block markers (such as ```json or ```):
[
  {
    "merged_question": XXXX,
    "sub_questions_id": [],
    "sub_questions": [],
    "task_type": XXXX,
    "reasoning": XXXX
  },
  ...
]

## Next, input a set of questions and corresponding answers:
[
  <question, answer>,
  <question, answer>,
  ...
]

## Please analyze the input content and generate as many results as possible according to the requirements while ensuring quality:
\end{promptbody}

\vspace{0.4em}
\noindent\textcolor{PromptMuted}{\small When answers are omitted, the input block uses question entries only; all other instructions remain unchanged.}

\subsection{Merged Question Solving}
\label{app:prompt-b3}

\begin{promptmeta}{B3 \textbar{} Merged Question Solving}
\begin{tabularx}{\linewidth}{@{}l X@{}}
\toprule
\textbf{Purpose} & Solve a synthesized merged question using sub-questions, sub-answers, and a reasoning trace. \\
\textbf{Input} & \texttt{merged\_question}, sub-questions with answers, and \texttt{reasoning} \\
\textbf{Output} & Runnable Python code (numerical tasks) or JSON with \texttt{thinking} and \texttt{answer} (logical tasks) \\
\bottomrule
\end{tabularx}
\end{promptmeta}

\begin{promptbody}
You are a knowledgeable and helpful financial domain expert. I now have a Question composed of at least 2 sub-questions. All sub-questions and their standard answers, as well as the overall approach for synthesizing this Question, are known.

## Your task
First read and understand the Question and Reasoning, and determine the type of the Question.
Questions are divided into two categories: numerical calculation and logical reasoning. The former aims to obtain answers by using relevant numerical values for calculation or comparison; the latter cannot rely solely on numerical calculation, but requires reasoning or summarization based on textual content to obtain answers.
Then, please combine the input sub-questions and sub-answers to generate a solution for the Question.
1. For numerical calculation Questions, generate runnable Python code to solve them. Code requirements: concise and clear, directly runnable, meaningful variable names, necessary input data definitions included, code comments and final output included.
2. For logical reasoning Questions, generate a clear step-by-step thinking process, then generate the answer, and return in the following plain-text JSON format. Do not include any code block markers (such as ```json or ```):
  {
    "thinking": generate thinking process in English,
    "answer": generate answer in English
  }

Below I provide this information to you. Please help me answer the Question as required.
## Input information
Question: [<full question text>]
SubQuestions_and_Answers:
[
  {subquestion: [<sub-question text>], answer: [<sub-question answer>]},
  {subquestion: [<sub-question text>], answer: [<sub-question answer>]},
  ...
]
Reasoning: [<approach for synthesizing the Question>]

## Please solve the Question as required. Note: do not question the correctness of the sub-question answers
\end{promptbody}

\subsection{Answer--Passage Consistency Checking}
\label{app:prompt-b4}

\begin{promptmeta}{B4 \textbar{} Answer--Passage Consistency Checking}
\begin{tabularx}{\linewidth}{@{}l X@{}}
\toprule
\textbf{Purpose} & Verify whether a manually annotated answer is supported by the relevant passage. \\
\textbf{Input} & \texttt{question}, \texttt{relevant\_passage}, \texttt{answer} \\
\textbf{Output} & JSON with \texttt{thinking} and \texttt{judgement} $\in$ \{Fully Consistent, Partially Consistent, Inconsistent\} \\
\bottomrule
\end{tabularx}
\end{promptmeta}

\begin{promptbody}
You are a knowledgeable and helpful financial domain expert. I now have a financial Question. Based on the Question, I found some relevant document passages (Relevant_passage) and manually annotated an answer (Answer). I now provide this information to you. Please help me check whether the Answer is consistent with the content described in Relevant_passage based on the Question.

## Your task:
Read and understand the Question, Relevant_passage, and Answer. Think about and judge whether the Answer conforms to Relevant_passage. First generate concise and clear judgment basis, then give a judgment conclusion (only one of 'Fully Consistent', 'Partially Consistent', or 'Inconsistent'). Return in the following plain-text JSON format. Do not include any code block markers (such as ```json or ```):
  {
    "thinking": generate thinking process in English,
    "judgement": give judgment result
  }

Below I provide the information to you. Please help me make the judgment as required.
## Input information:
Question: [<question>]
Relevant_passage:
[<relevant_passage>]
Answer: [<answer>]

## Please complete the judgment task as required:
\end{promptbody}

\makeatother
\endgroup

\end{document}